\documentclass[acmsmall]{acmart}
\usepackage{multirow}
\usepackage{amsmath}

\AtBeginDocument{%
  \providecommand\BibTeX{{%
    \normalfont B\kern-0.5em{\scshape i\kern-0.25em b}\kern-0.8em\TeX}}}

\setcopyright{acmlicensed}
\acmJournal{PACMHCI}
\acmYear{2023} \acmVolume{7} \acmNumber{CSCW2} \acmArticle{347} \acmMonth{10} \acmPrice{15.00}\acmDOI{10.1145/3610196}

\begin{document}

\title[MeetScript: Designing Transcript-based Interactions to Support Active Participation in Group Video Meetings]{MeetScript: Designing Transcript-based Interactions to Support Active Participation in Group Video Meetings}

\author{Xinyue Chen}
\email{xinyuech@umich.edu}
\affiliation{%
  \institution{University of Michigan}
  \state{Michigan}
  \country{USA}
}

\author{Shuo Li}
\email{shuolii@umich.edu}
\affiliation{%
  \institution{University of Michigan}
  \state{Michigan}
  \country{USA}
}

\author{Shipeng Liu}
\email{shipengl@usc.edu}
\affiliation{%
  \institution{Universtiy of Southern California}
  \state{California}
  \country{USA}
}

\author{Robin Fowler}
\email{rootsr@umich.edu}
\affiliation{%
  \institution{University of Michigan}
  \state{Michigan}
  \country{USA}
}

\author{Xu Wang}
\email{xwanghci@umich.edu}
\affiliation{%
  \institution{University of Michigan}
  \state{Michigan}
  \country{USA}
}
\renewcommand{\shortauthors}{Xinyue Chen et al.}

\begin{abstract}
While videoconferencing is prevalent, concurrent participation channels are limited. People experience challenges keeping up with the discussion, and misunderstanding frequently occurs. Through a formative study, we probed into the design space of providing real-time transcripts as an extra communication space for video meeting attendees. We then present MeetScript, a system that provides parallel participation channels through real-time interactive transcripts. MeetScript visualizes the discussion through a chat-alike interface and allows meeting attendees to make real-time collaborative annotations. Over time, MeetScript gradually hides extraneous content to retain the most essential information on the transcript, with the goal of reducing the cognitive load required on users to process the information in real time. In an experiment with 80 users in 22 teams, we compared MeetScript with two baseline conditions where participants used Zoom alone (business-as-usual), or Zoom with an adds-on transcription service (Otter.ai). We found that MeetScript significantly enhanced people's non-verbal participation and recollection of their teams' decision-making processes compared to the baselines. Users liked that MeetScript allowed them to easily navigate the transcript and contextualize feedback and new ideas with existing ones. 
\end{abstract}




\maketitle

\begin{figure}[h]
\includegraphics[width=0.96\textwidth]{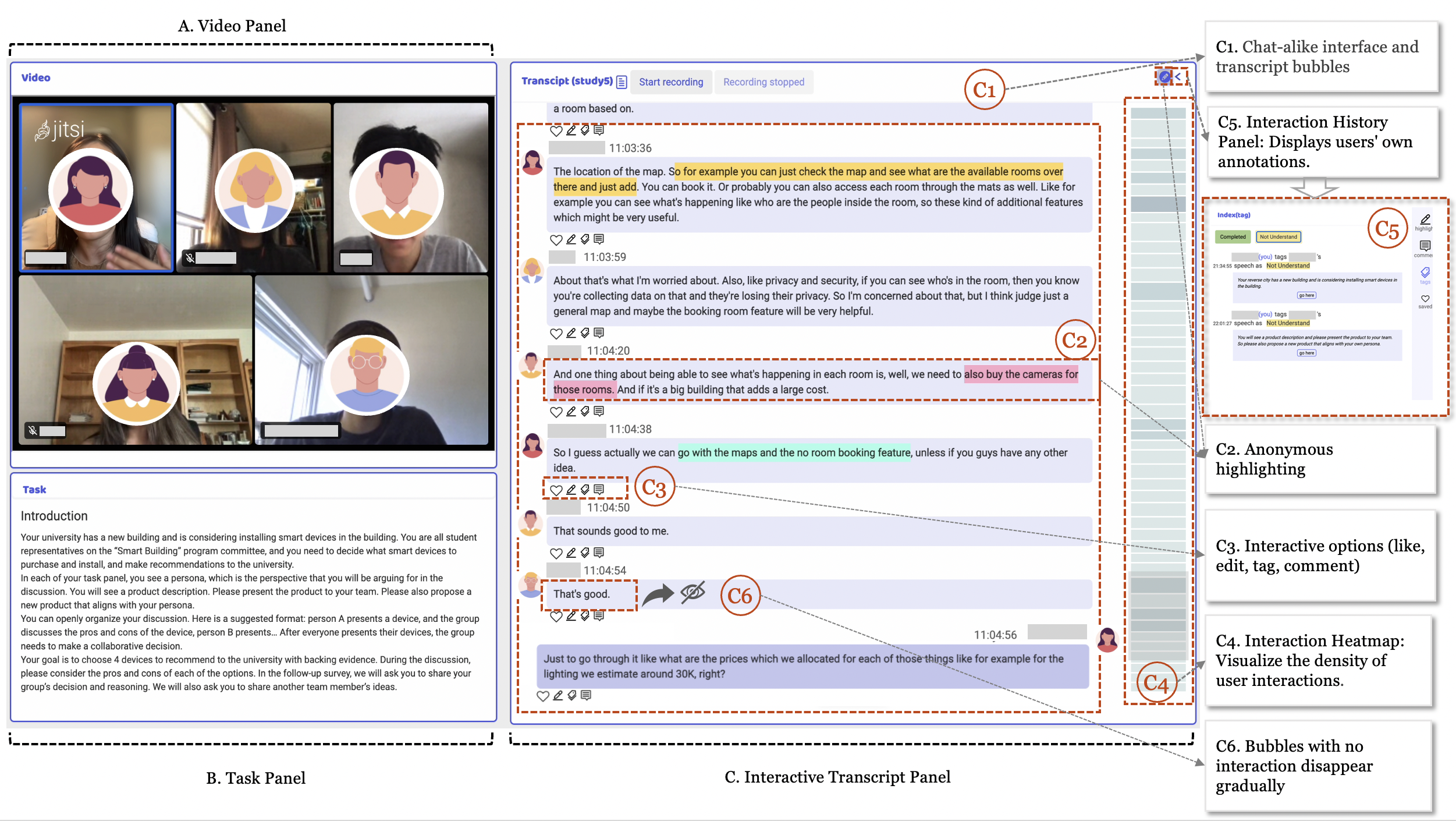} 
\caption{\textbf{MeetScript User Interface.} MeetScript supports group video meetings through (A) a Video Panel and (C) an Interactive Transcript Panel. Users' conversations are transcribed in real-time and displayed in a chat-alike interface through transcript bubbles (C1). Users have a suite of options to interact and contribute to the conversation unobtrusively through the transcript. Users can highlight (C2), like, edit, tag, and comment on each transcript bubble (C3). To help users process the most essential information, transcript bubbles without user interaction gradually disappear over time (C6). Users can navigate the transcript using the Interaction Heatmap (C4) and Interaction History Panel (C5).}
\label{Fig.system}
\end{figure}

\section{Introduction}

The demand for remote work and online meetings within and between distributed teams has significantly grown in recent years. Task-oriented online meetings on complex group projects happen every day over videoconferencing platforms \cite{cao2021large,larsson2020communication}. For example, Zoom has 300 million daily meeting participants as of March 2021, a 2900\% increase from December 2019 \cite{Zoomusers}.

With the boom of videoconferencing in both workspace and educational settings, it is critical to identify opportunities on how to best support video meeting experiences. Prior work has shown that video meetings introduce challenges in participation, transparency, and efficiency for several reasons. 
First, video meetings exacerbate the problems with active participation in group discussions. Although prior work has shown that active participation from attendants is essential to make a discussion effective \cite{cook2009building, lunenburg2010communication}, reasons such as information loss and conformance pressure prevent people from actively contributing \cite{green1991today, heath2013human}. Second, turn-taking and back-channeling become more challenging in video meetings. The lack of physical proximity and the absence of social cues in video meetings make it difficult for attendees to interrupt the main communication channel to contribute to the discussion \cite{karl2021virtual, chang2016challenges, dhawan2021videoconferencing, wainfan2004challenges, larsson2020communication}. 
Third, since multi-tasking is frequent \cite{cao2021large}, and attention span is limited \cite{bailey2006need, kuzminykh2020low}, online meeting attendants often miss content, get lost in long monologues and struggle to keep up with the conversations \cite{wainfan2004challenges}.

To address the challenges around active participation and keeping up with long conversations, one line of work explores post-meeting support, e.g., dashboards that provide summaries or visualizations of group dynamics to help members reflect on their meeting performance \cite{chandrasegaran2019talktraces, ullmann2019visualisation, samrose2021meetingcoach, samrose2018coco}. Other work provides in-situ support through augmenting meeting attendants' facial expressions \cite{murali2021affectivespotlight, cho2021want, he2021you} and providing parallel chat \cite{sarkar2021promise}. However, parallel chat messages in video meetings can be distracting since they are disconnected from the discussion context \cite{sarkar2021promise}. 
Showing transcripts and captions in real-time is beneficial for understanding educational videos \cite{kim2014data, fang2021notecostruct} and verbal conversations e.g., in cross-cultural understanding \cite{gao2014effects} and meetings involving people with disabilities \cite{kafle2021deaf}.
\textit{However, it remains an open question} whether real-time transcripts can be an effective communication space that provides parallel communication channels for meeting attendants. 

In text-based communication, e.g., discussion forums and chats, researchers have explored collaborative sense-making techniques, such as inserting comments and tags \cite{zhang2018making} to help discussants keep up with the long textual content. 
In this work, we build off prior research on structuring and visualizing documents \cite{ruan2019bookbuddy} and collaborative annotation techniques for making sense of text-based chats \cite{zhang2018making, geller2020confused, pan2017task}. We consider how to apply these techniques in situ to enrich video meeting experiences. We explore ways meeting participants can view real-time transcripts in a structured way, interact with the transcripts to insert new ideas, mark up important content, express confusion, and keep up with the conversation through a collaborative tangible document. We also explore ways to provide essential information to users while not overwhelming them.

From a speed-dating study with 22 participants, we validated the user needs that many interviewees reported a lack of channels to express ideas, show confusion and connect new topics with existing ones. Participants also found it challenging to manage to speak and listen at the same time and keep track of conversational threads. We probed into participants' preferences and concerns about facilitating video meetings with real-time transcripts as a parallel participation space. We found that participants liked the idea of using the transcripts as a way of participation. They wanted diverse interactions that adapted to their situational needs. They emphasized that such transcripts should not place an extra cognitive load on them and that the readability of the transcript matters to them.

Based on the findings, we iteratively develop \textit{MeetScript}, a system that leverages real-time transcripts to support active participation in group video meetings (Figure ~\ref{Fig.system}). MeetScript provides a real-time transcript interface, visualizing the conversation through a chat-alike interface to ensure that the transcript is readable and aligns with users' conversation flow. MeetScript provides a suite of interactions for users to collaboratively like, highlight, comment, tag, and edit each transcript bubble. As the meeting progresses, MeetScript gradually hides extraneous information to keep essential information on the transcript, with the goal of reducing the cognitive load required on users to process the information in real-time. Finally, MeetScript visualizes the participation density in an "Interaction Heatmap" and an "Interaction History Panel".

To evaluate the effectiveness of MeetScript on enhancing meeting participation and understanding, we conducted a between-subjects experiment with 80 users in 22 teams. Participants were randomly assigned to three conditions, including MeetScript and two baselines. The first baseline is a business-as-usual meeting setup using Zoom with optional note-taking software such as Google Docs. The second baseline adopts a widely used live transcription service, Otter.ai \cite{otter}, as an add-on to Zoom. Otter.ai supports basic collaborative annotations on a live transcript. 
We found that users in MeetScript demonstrated significantly more non-verbal participation through the transcript compared to the Zoom (p=0.005) and Zoom+Otter.ai conditions (p=0.007). Importantly, MeetScript users regarded their team members' transcript-based interactions as valuable contributions to the discussion, whereas Zoom+Otter.ai users did not. MeetScript users showed better recollection of their team's decision-making processes and individual team members' ideas than the two baselines. MeetScript users displayed entirely different sentiments towards using transcripts to facilitate video meetings in contrast to Zoom+Otter.ai users. Specifically, MeetScript users considered the transcripts to be more helpful in expressing their opinions, understanding their teammates' ideas, increasing the transparency of the conversation, resolving misunderstandings, and reviewing and synthesizing previous discussions compared to the Zoom+Otter.ai users.

We consider the design and development of MeetScript to be a first step in extending the communication space to support active participation and understanding in video meetings. We provide both quantitative and qualitative evidence that live transcripts can be used as an extra communication space when the system visualizes transcripts through a readable interface that aligns with the conversation flow, invites contextualized and low-effort user interactions, and embeds information filtering mechanisms that help users locate important information.

\section{Related Work}

We review previous research on the challenges around active participation in group meetings, especially video meetings. We summarize existing tools that support active participation in virtual meetings, tools to support back-channeling and transcript-based interactions, and collaborative sense-making techniques to support text-based communication. 

\subsection{Challenges on Active Participation in Physical and Video Group Meetings}
\subsubsection{Participation Challenges in Group Meetings}
Group meetings are events where participants discuss, negotiate, present, and create materials together in a communicative manner \cite{jones1999silence, straus1996getting}. To be effective in these meetings, it is important for participants to actively participate \cite{masek2021defining}, including expressing ideas and asking questions \cite{straus1996getting}. 
However, prior work has summarized the challenges that prevent people from actively participating. First, marginalized groups may face difficulties in engaging in the discussion \cite{fowler2014talking,jones1999silence}. Second, people lack channels to contribute without interrupting the flow of the conversation \cite{kirshenbaum2021traces}. Third, active participation requires adequate information exchange and sufficient understanding among participants \cite{robertson2005active}. This can be hard to achieve when participants have difficulty keeping up with all the information being exchanged in the discussion. As a result, people may be unable to participate and contribute effectively \cite{beers2006common}. 

\subsubsection{Participation Challenges Specific to Group Video Meetings}
Online video meeting environments exacerbate many of the problems identified above \cite{myers2008fundamentals,larsson2020communication, fahy2006online, wohn2017face, jonassen2001communication}.
For example, interlocutors cannot fully observe their partner's physical environment \cite{wainfan2004challenges}, and non-verbal cues are often absent, making it difficult for meeting participants to intervene \cite{dong2012one, seuren2021whose}, and talking-over-each-other is more likely to happen \cite{seuren2021whose}.   
Additionally, misunderstandings may occur \cite{jonassen2001communication, cho2021want, tu2018you}, since there are not sufficient communication channels for attendees to clarify concepts compared to in-person meetings \cite{chang2016challenges}. 
Recent research shows that compared to in-person meetings, people in videoconferencing are more likely to miss parts of a meeting because of interruptions, distractions, and multitasking \cite{cao2021large, suh2018s, junuzovic2011did}. 
In summary, in group video meetings, people face more difficulties in participating actively, observing turn-taking rules, building mutual understanding, and keeping up with the conversation.

\subsection{Tools to Increase Participation in Online Video Meetings}

\subsubsection{Tools to Support Participation Awareness Post Meetings}
One line of work focuses on providing post-meeting support through dashboards and visualizations to increase users' awareness of their engagement levels\cite{samrose2018coco,samrose2021meetingcoach}. For example, Coco visualizes user attention, participation, and speech overlaps after a meeting to raise user awareness of conversational dynamics \cite{samrose2018coco}. MeetingCoach provides a post-meeting dashboard that shows engagement, tone, and speaking patterns \cite{samrose2021meetingcoach}. However, these tools only help users understand their participation levels and patterns after meetings and do not assist with real-time participation. 

\subsubsection{Tools to Support In-situ Participation}
Tools are also designed to help people keep up with the conversation in situ. One line of studies supports participation in video meetings by simulating the face-to-face meeting experience by amplifying expressions and body language, as well as strengthening visual cues \cite{namikawa2021emojicam, murali2021affectivespotlight}. For instance, MeetingCues uses emojis and emotion visualization to support situational awareness and reflection \cite{aseniero2020meetcues}. Existing commercial videoconferencing platforms use emojis, raising hands, and polls to support non-verbal participation from meeting attendants. However, research has shown that such interaction mechanisms can be delayed, are not contextualized in the conversation, and thus less useful \cite{karl2021virtual, wiederhold2020connecting, cho2021want, kohnke2020facilitating}. Other work aims to improve participation by increasing participants' group awareness in real-time \cite{ chandrasegaran2019talktraces, junuzovic2011did}. For example, TalkTraces identifies emerging discussion topics, generates future meeting agendas, and visualizes group dynamics. As a result, attendants are motivated to participate and engage more in the discussion \cite{chandrasegaran2019talktraces}. 

\subsubsection{Back-channeling and Concurrent Feedback}
In computer-mediated communication, prior work has emphasized the importance of enabling meeting attendants to offer immediate feedback to each other, which is also referred to as concurrent feedback or "back-channel" feedback\cite{dennis1998testing}.
Concurrent feedback refers to nonverbal acts used by attendees to show that they are actively listening and participating in the discussion, such as nodding or making non-lexical utterances \cite{dennis1998testing}. 
Almost all commonly used video meeting platforms, e.g., Zoom, and Google Meet provide parallel chat features \cite{bin2020Zoom}, which enable questions, clarifications, posting resources, etc \cite{berry2019role}. However, prior work shows that parallel chats can easily cause distraction and off-topic discussions \cite{sarkar2021promise} because they lack references to the audio and video context \cite{sarkar2021promise}. Contextualized feedback has been explored in a live presentation setting. For example, PeerPresents enables students to offer and receive feedback synchronously during presentations  \cite{shannon2016peerpresents, warner2023slidespecs}. These studies emphasized the importance of providing contextualized feedback to make the feedback more useful, which is also suggested in learning sciences literature with regard to the properties that make feedback effective \cite{nelson2009nature}.

In summary, prior work aiming to provide in-situ participation support in video meetings does not focus on extending the communication space for meeting attendants, and existing parallel channels in video meetings do not contextualize users' chat messages with existing discussions. Prior studies on feedback show that contextualized and localized feedback is more beneficial for synchronous conversations such as live presentations.

\subsection{The Promise of Using Transcripts As an Extra Communication Space in Video Meetings}
One challenge that hinders people from participating actively in meetings is that they can not keep up with the discussion \cite{robertson2005active}. One common method of keeping up with discussions in online meetings is note-taking. However, taking deliberate notes during meetings can be cognitively challenging and distracting \cite{piolat2005cognitive}. Research has shown that users can be overwhelmed by note-taking and do not have the cognitive capacity to think critically and encode the ideas being exchanged during a discussion \cite{fang2022understanding, fanguy2021collaboration}. 

Providing users with transcripts and captions in real-time has been found beneficial for them to understand verbal conversations or video content. For example, live transcripts were found to benefit cross-cultural understanding \cite{gao2014effects, pan2017task}, and educational video comprehension \cite{grgurovic2007help}. However, at the same time, prior work has shown that searching for information and navigating a long transcript document can be difficult, especially during real-time conversations \cite{pavel2014video, pavel2015sceneskim, kim2014data, huber2019b}. Most existing videoconferencing platforms provide real-time transcripts to enhance accessibility. Apart from built-in transcription services, add-on transcription tools have the potential to provide an extra communication space for users to annotate content and assign tasks. For example, Otter.ai \cite{otter} is a widely adopted transcription service connected to Zoom. It provides a live transcript with annotation features, enabling users to mark important content, as shown in Figure ~ \ref{Fig.otter}. However, research showed that people mostly used this for post-meeting transcription and review purposes \cite{chiaonline}. It remains an open question whether live transcripts can be an effective communication space that provides in-situ parallel communication channels for meeting attendants.

\subsection{Information Organization and Collaborative Sense-making for Textual Interactions} 
Prior research has suggested ways to visualize long and verbose documents in a more readable fashion, including chunking long documents into smaller pieces \cite{miniukovich2019guideline}, and increasing the interactivity of the reading interface \cite{ruan2019bookbuddy}. More recent work explored social annotation methods to help groups of people collaboratively make sense of a piece of content, e.g., enabling tagging in online forums \cite{geller2020confused}.
In text-based communication, e.g., real-time chats, researchers have explored a variety of collaborative sense-making techniques to help people refer back and forth in long conversations \cite{zhang2017wikum, zhang2018making, tian2021system}. One notable system is Tilda, which allows users to tag chat messages in Slack as a way to collaboratively summarize and make sense of the conversation\cite{zhang2018making}. Other studies also showed the benefits of visualization techniques in helping people navigate textual conversations. For example, ConTovi used a text-mining technique to visualize topic change \cite{el2016contovi}, and T-cal disentangled interleaving conversations \cite{fu2018t}.

In addition to text-based communication, researchers have looked into methods that enable users to navigate or process transcripts easily. For example, researchers explored having learners collaboratively annotate the transcripts of educational videos and found that highlighted keywords in transcripts supported learners' comprehension \cite{torre2022video, fang2021notecostruct, fanguy2021collaboration, kam2005livenotes}. Researchers have also explored methods to present reduced information from a transcript, e.g., summaries or keywords in contrast to the full transcript, but found that the reduced version may introduce new misunderstandings \cite{shi2018meetingvis}. Previous work pointed out that providing more visual cues in the textual transcript can reduce the cognitive load on transcript navigation \cite{yadav2015content, kim2014data}.

\textbf{As a Summary of existing work, }research has shown that problems in video meetings persist. Participation channels are limited, concurrent feedback is often missing, and it can be challenging for people to take turns and keep up with the heightened discussion effectively. Existing systems have not done a satisfying job of providing group members enough opportunities and channels to engage in the conversation and keep up with and understand the discussion in real-time. Existing back-channeling interfaces do not allow meeting attendants to contextualize messages or help them make sense of the information. Although live transcripts are mundane in most videoconferencing platforms now, they are mainly designed to enhance accessibility and for post-meeting reviews. It remains an open question whether live transcripts can be an effective communication space that provides in-situ parallel communication channels for meeting attendants. Inspired by prior work on collaborative sense-making techniques in text-based communication, in this work, we explore ways to enrich video meetings with interactive real-time transcripts and, at the same time, techniques to reduce the cognitive load required on users to process such parallel information in real time.

\section{Formative study}
As shown in prior work, live transcripts can help people catch up with the content in video meetings. However, dense transcripts may present an extra cognitive load on users when they are engaged in a conversation. The goal of the formative study is to 1) validate user needs around active participation and understanding and 2) probe into users' preferences and concerns on using transcripts to facilitate video meetings.

\subsection{Method}
\subsubsection{Design Concepts and Storyboards}
\begin{figure}[h!] 
\centering 
\includegraphics[width=0.8\textwidth]{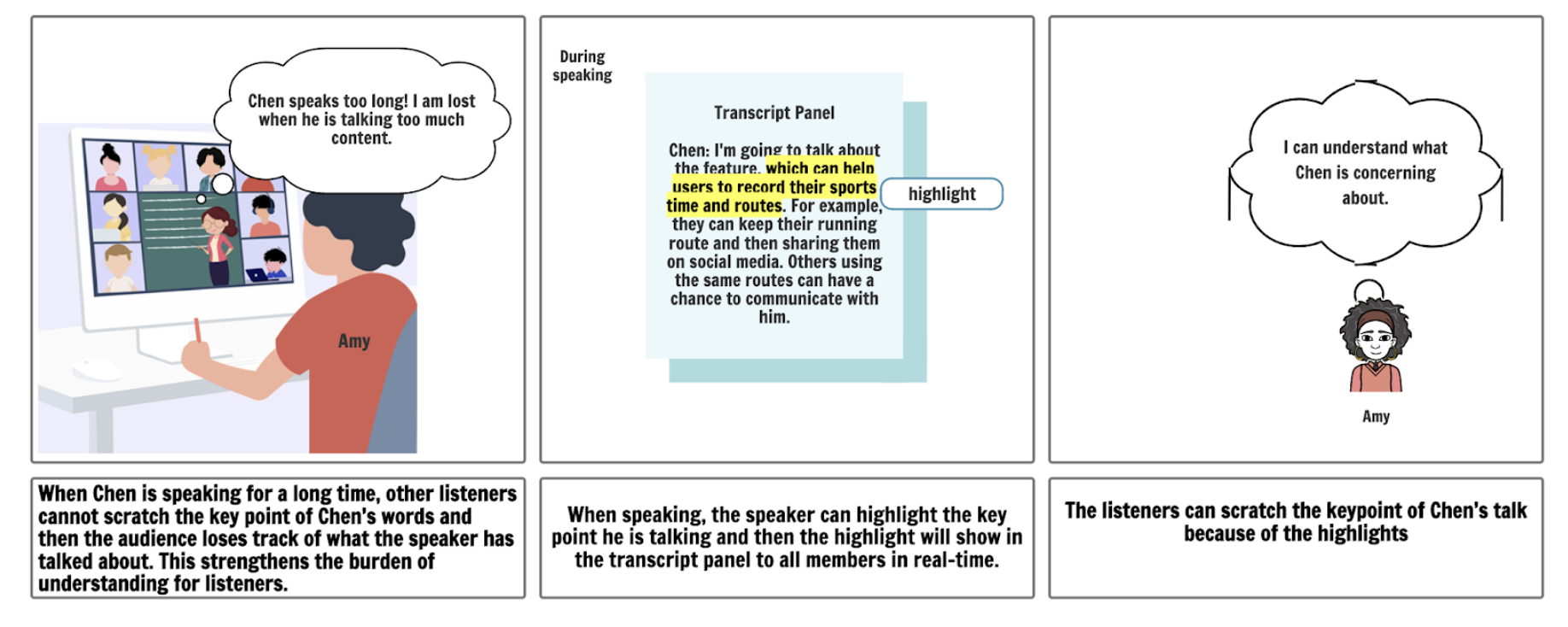} 
\caption{\textbf{Storyboard 10: Highlight what is important in the live transcript when speaking.} Participants thought this could help them keep up with the conversation and quickly get the key ideas.
}
\label{Fig.sample}
\end{figure}

Although live transcripts are mundane in most videoconferencing platforms, they are mainly designed to enhance accessibility and support post-meeting reviews. It remains an open question of how to design real-time transcripts to make them effective parallel participation channels for video meetings without distracting or overwhelming users. In this study, we designed scenarios using storyboards to explore when and how meeting attendants may want to use transcripts. 
We consider storyboards to be a useful tool in probing the design space for videoconferencing tools since they help participants visualize the scenario quickly. 
We iteratively improved the storyboards to ensure that they neutrally reflect the design concepts and can stimulate discussion among participants\cite{luria2020social}. 
10 storyboards were used in the formative study.
Each storyboard describes a problem scenario that occurs in a video meeting, possible design solutions, and hypothetical outcomes. 
Figure ~ \ref{Fig.sample} shows an example. We initially created 20 storyboards and narrowed them down to 10 by grouping similar ideas. The ten storyboards specifically focused on how users might interact with transcripts during synchronous meetings, and many of the interaction mechanisms were inspired by prior work on document visualization  \cite{lampe2007follow, miniukovich2019guideline} and social annotation tools \cite{zhang2018making}. The whole list of storyboards used can be found in the supplementary materials. A brief version is shown in Table \ref{tab:concept-storyboard} in the Appendix.

\subsubsection{Procedures}

\begin{figure}[h!] 
\centering 
\includegraphics[width=0.9\textwidth]{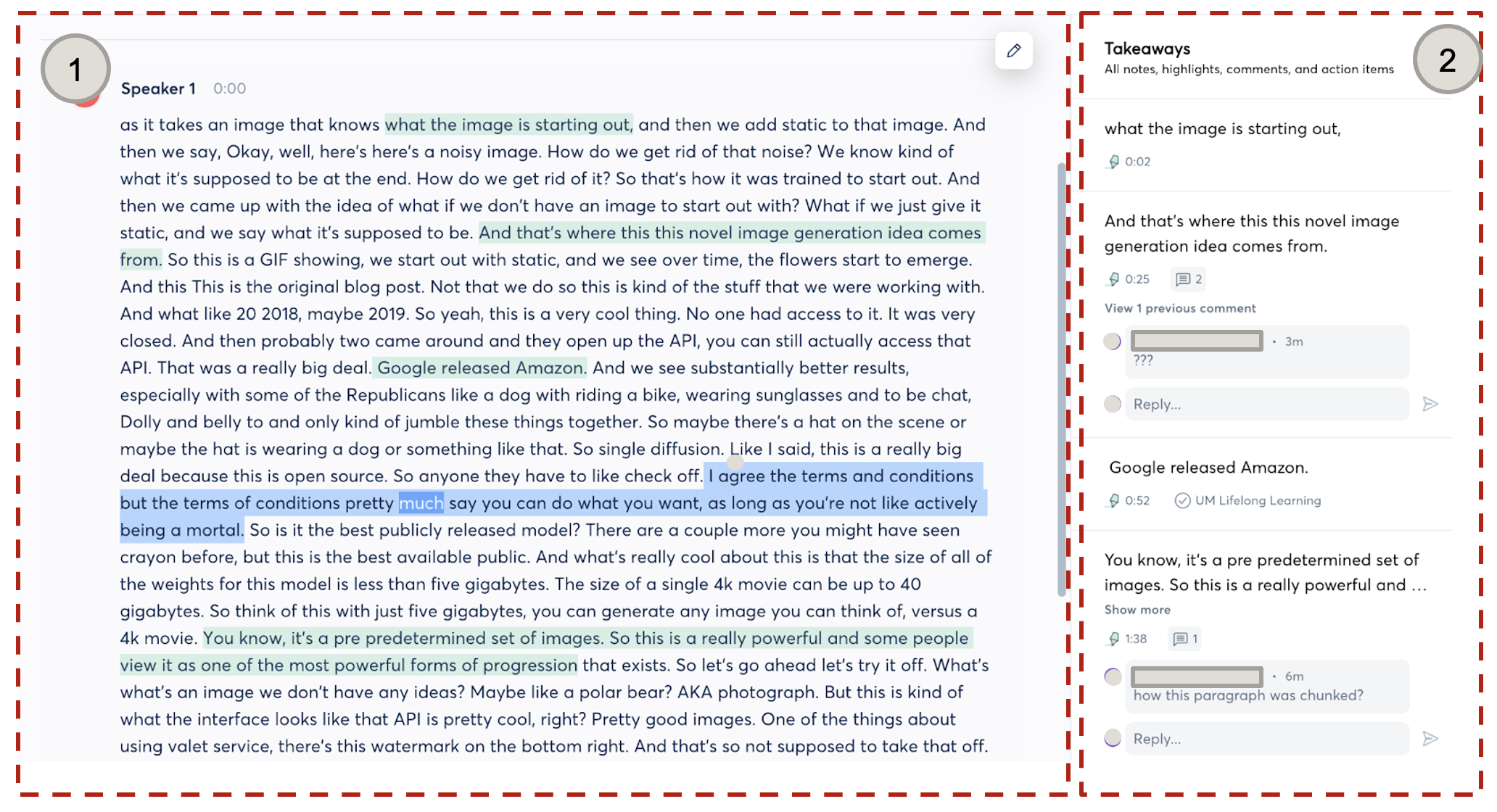} 
\caption{\textbf{Otter.ai Interface.} (1) It provides a live transcript of a Zoom meeting. Users can highlight the document. (2) A takeaway panel integrates user highlights. Users can add comments to the highlights they marked in this panel.
}
\label{Fig.otter}
\end{figure}

We recruited participants through social media, including Reddit and Twitter. Participants were asked to fill out a short screening survey about their past online meeting experiences. There were 30 valid responses, and we scheduled the speed-dating study with 22 participants.  The formative study was conducted (13 males, 9 females, average age at 28) over Zoom. All of the participants have used Zoom extensively. The demographic information is shown in Table. \ref{demo}.  In each session, we first showed participants the interface of Otter.ai \cite{otter}, which is a widely used transcription service. Otter.ai provides real-time transcripts with basic annotation and note-taking features, as shown in Figure. \ref{Fig.otter}. The rationale of showing participants Otter.ai is to give them a concrete idea of what might a real-time interactive transcript panel look like and how they may be able to interact with it. This makes it easier for the participants to read and evaluate our design concepts subsequently. 

A researcher then presented the storyboards in a random order through screen sharing. For each storyboard, the researcher introduced the scenario and the solution and gave participants time to read.
Participants were then asked to share their thoughts on each solution. The speed-dating interviews lasted for an average of 75.6 minutes. The study is IRB approved, and each participant received 25 dollars as compensation. The interviews were transcribed and two researchers used the 
Affinity Diagram approach \cite{ lucero2015using} to analyze the data. In the analysis, two researchers rearranged all quotes iteratively based on emerging affinity to one another through communication and critique. We grouped users' feedback, including their preferences and concerns, on each design solution as shown in the storyboards and narrowed down the design space.

\subsection{Findings}

We found participants interested in using transcript-based interactions to insert ideas and give feedback. We found making the interaction anonymous could help some users. Besides, the study revealed participants' concerns with high cognitive load in using transcripts.

\subsubsection{Transcripts can help me keep up with the conversation}
Almost all interviewees shared that real-time transcripts could help them keep up with the conversation. U13 said, \textit{"You know, before I speak, I cannot pay attention to what others are talking about. But if I can review what others talked about through transcripts. That's great."}. Besides, people thought it would help them focus on the discussion without stressing about missing some content. U22 said, \textit{"Since I already pinned the whole transcript, I only need to take down the most salient points.}

\subsubsection{Transcripts-based interactions may help me participate concurrently}
Participants described scenarios when they wanted to use transcript-based interactions. 

\textbf{Transcript-based interaction as a way of parallel communication.}
Participants thought transcript-based interactions could be an effective parallel communication channel to signal the speaker that they were paying attention. As U1 mentioned, \textit{"In online meetings, only one person can speak at a time. Then what you showed in the storyboard, you can react on the transcript using emojis, like, and dislike, which can be a way to communicate attitudes. You don't need to unmute yourself. It's more convenient. ''} Some people thought it is user-friendly to less-vocal people, as U15 said, \textit{"As someone who doesn't want to speak up in a group discussion, I will be more comfortable interacting with the transcript. I always feel difficult to find an opportunity to speak.''}

\textbf{Show confusion and ask questions through transcript}
In the storyboards, we propose designs where users can annotate the transcript in real-time to show confusion. Participants liked to have more ways to express confusion, as U4 said,\textit{ "I like the way of highlighting my confusing part in the transcript to show that I don't understand it. "} Participants liked using transcript to contextualize their questions, as U18 said, \textit{"I can imagine there is a space to connect chat messages with the transcript. People usually send questions in the chat. And like what you show, you can add a question below a turn on the transcript. Then everyone knows what your question refers to. ''}

\textbf{Provide anonymous channels for participation.}
Many participants mentioned that they wanted to have more anonymity during the discussion. U9 said, \textit{"I like the anonymous comment feature you mentioned. I think anonymity is a basic feature, that is, in all interactions except speaking, users can choose to be anonymous or not." } Providing anonymous communication spaces could be especially useful for certain groups of users, as U7 noted, \textit{"I am hesitant to express my confusion during small group project meetings because I am worried that other students will think I am not intelligent enough to do well in the course. As the only non-native speaker in the group, I feel embarrassed to ask questions. However, if the option to ask anonymously was available, I would certainly seek clarification."}

\textbf{Provide diverse interaction methods on the transcript.} It is important to note that participants expressed a need for diverse ways of interacting with the transcript, as U12 said, \textit{"It's hard to say I prefer which interaction. I found different interactions can target different needs. I can easily highlight something to show what I am paying attention to, while I can assign a tag to a transcript message to summarize the conversation when needed. It depends on what kind of meetings we have. "} This suggests that offering a variety of interaction options may improve user experience.

\subsubsection{The readability of the transcript will influence my use.}
When showing people the Otter.ai interface, participants mentioned their concerns that it would add extra cognitive load to them to read the full transcript. Besides, they worried about the accuracy of the transcript. Many participants emphasized the challenges of reading the full transcript, as U2 mentioned, \textit{"I don't usually open the transcript in Zoom when I have meetings with others. I don't think I have time to read the transcript during the discussion. And you know that finding some specific information in the long and less organized transcript is time-consuming."} 
Some participants thought simplifying the information matters, as U22 mentioned, \textit{"Only useful information kept in the transcript is enough. I do not need to review all the transcripts. "}

\subsection{Summary of Design Goals}
Based on our formative study and prior research, we summarized the design requirement when using real-time transcripts to enrich video meeting experiences:
\begin{itemize}
\item Make it easier for users to navigate and locate information on the transcripts.
\item Present the essential information on the transcript without adding extra cognitive load on the users.
\item Provide a variety of interaction methods with the transcript so that users have a choice. 
\item Enable people to participate anonymously. 
\end{itemize}

\section{MeetScript: Enhancing Video Meeting Participation and Understanding through an Interactive Real-Time Transcript}

Based on the design requirements, we developed MeetScript, a web-based system that provides transcript-based interactions to support active participation and understanding in video meetings.
An overview of the MeetScript system is shown in Figure.\ref{Fig.system}. 

MeetScript's front-end interface has three components: 1) a video panel similar to traditional videoconferencing platforms; 2) a task panel for users to upload meeting agendas; 3) an interactive transcript panel. The interactive transcript panel provides a real-time transcription of users' speeches, powered by the Microsoft Azure Speech-to-text API \cite{speech}. 
Different from live transcription in existing videoconferencing platforms, MeetScript visualizes the discussion in a chat-alike interface, differentiating users' own messages from other attendants' messages. 
MeetScript supports a suite of interactions on each transcript bubble (Figure~\ref{Fig.edit}, including edit (A), tag(B), comment(C), and highlight(D), like (E). The transcript bubbles enable users to contextualize their feedback and comments, addressing the challenges of concurrent communication identified in prior work where chat messages were disconnected from the discussion \cite{sarkar2021promise}. 
To satisfy the design requirement of not overloading users, MeetScript implements two mechanisms to keep essential information on the transcript. First, transcript bubbles without any user interaction will disappear after 3 minutes. Second, MeetScript provides an Interaction Heatmap and an Interaction History Panel for users to locate information more easily. 
The design of MeetScript went through several rounds of pilot testing. The goal is to ensure the transcript is interactive, readable, and reviewable. We will describe the detailed design and implementation below. 

\subsection{Visualize Transcript through a Chat-alike Interface} To reduce the cognitive load for users to read the transcript during a conversation, we chunked the transcript into short and readable pieces. Using the Azure Speech-to-text service \cite{speech}, MeetScript identifies a single utterance in one's speech by the duration of silence and then puts every single utterance into a transcript bubble. Transcript bubbles are visualized in a way that resembles a commonly used chat interface. MeetScript recognizes speakers by the user token in their corresponding web browsers. On the one hand, it differentiates from existing transcription services by providing a clear structure of the document; On the other hand, the transcript bubbles become the basic unit for collaborative interactions, which enables contextualized feedback.

\subsection{Contextualize Concurrent Feedback Through Transcript-based Interactions}
\begin{figure}[t!] 
\centering 
\includegraphics[width=0.9\textwidth]{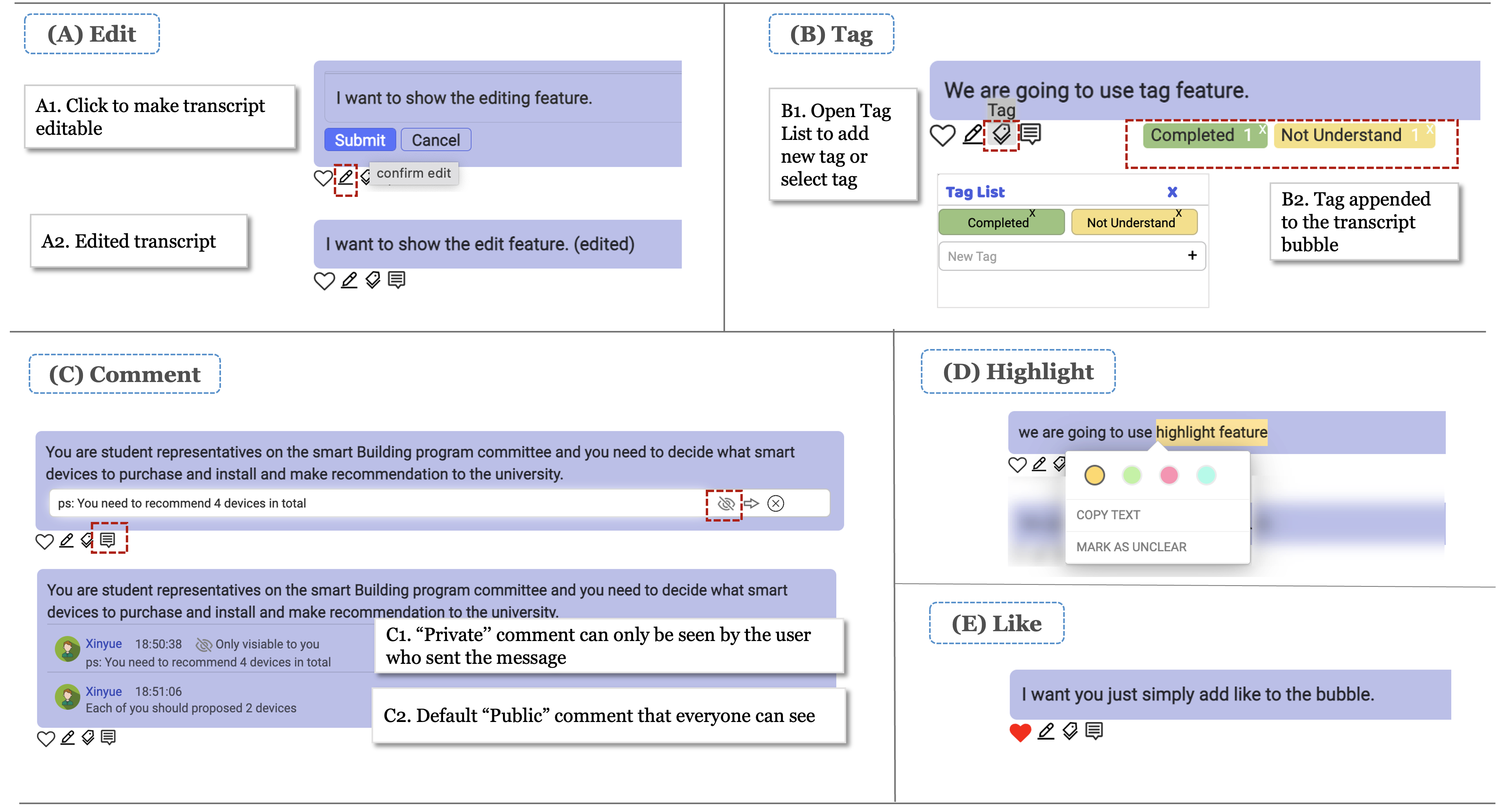} 
\caption{\textbf{A suite of interaction methods on transcript bubbles.} \textit{(A) Edit.} Users can add/edit/delete content. \textit{(B) Tag.} Users can define and assign tags. \textit{(C) Comment.} Users can choose to send comment publicly (C1) or privately (C2). \textit{(D) Highlight.} Users can highlight texts in different colors. \textit{(E) Like.} Users can use "like".}
\label{Fig.edit}
\end{figure}

MeetScript provides a suite of five interaction options, as shown in Figure~\ref{Fig.edit}. The five options are inspired by prior work on collaborative sense-making in text-based communication \cite{zhang2018making, tian2021system, fang2021notecostruct}. We consider different options to enable users to engage at different effort levels during a conversation. First, MeetScript enables users to "edit" a transcript bubble when they catch mistakes, as shown in Figure~\ref{Fig.edit}(A). To help users extract high-level information from transcripts, users can create customized tags and append a tag to a transcript bubble (Figure~\ref{Fig.edit}(B)). Tags help people summarize ideas from the conversation. One tag can be repeatedly added to a bubble, and the interface will show a count. To help users insert new ideas and ask and answer questions, MeetScript supports contextualized commenting (Figure~\ref{Fig.edit}(C)), which addresses the problem where chat messages and verbal conversation are disconnected in existing parallel communication channels of videoconferencing. Users' comments are public by default (Figure ~ \ref{Fig.edit}(C1)), whereas they can choose to make a comment private (Figure ~ \ref{Fig.edit}(C2)). 
MeetScript offers two interactions that require relatively lower effort levels, "highlight" (D) and "like" (E). 
Users can choose different colors in highlighting. 
All user interactions are seen by each other in real-time. The "edit", "tag", "highlight", and "like" interactions are all anonymous by default.

\subsection{Techniques to Keep Essential Information and Reduce Cognitive Load}
Live transcripts may add extra cognitive load to users. In MeetScript, we designed mechanisms to help users keep the most useful information.

\subsubsection{Gradually Delete Transcript Bubbles without Interactions.} 
\begin{figure}[t!] 
\centering 
\includegraphics[width=0.8\textwidth]{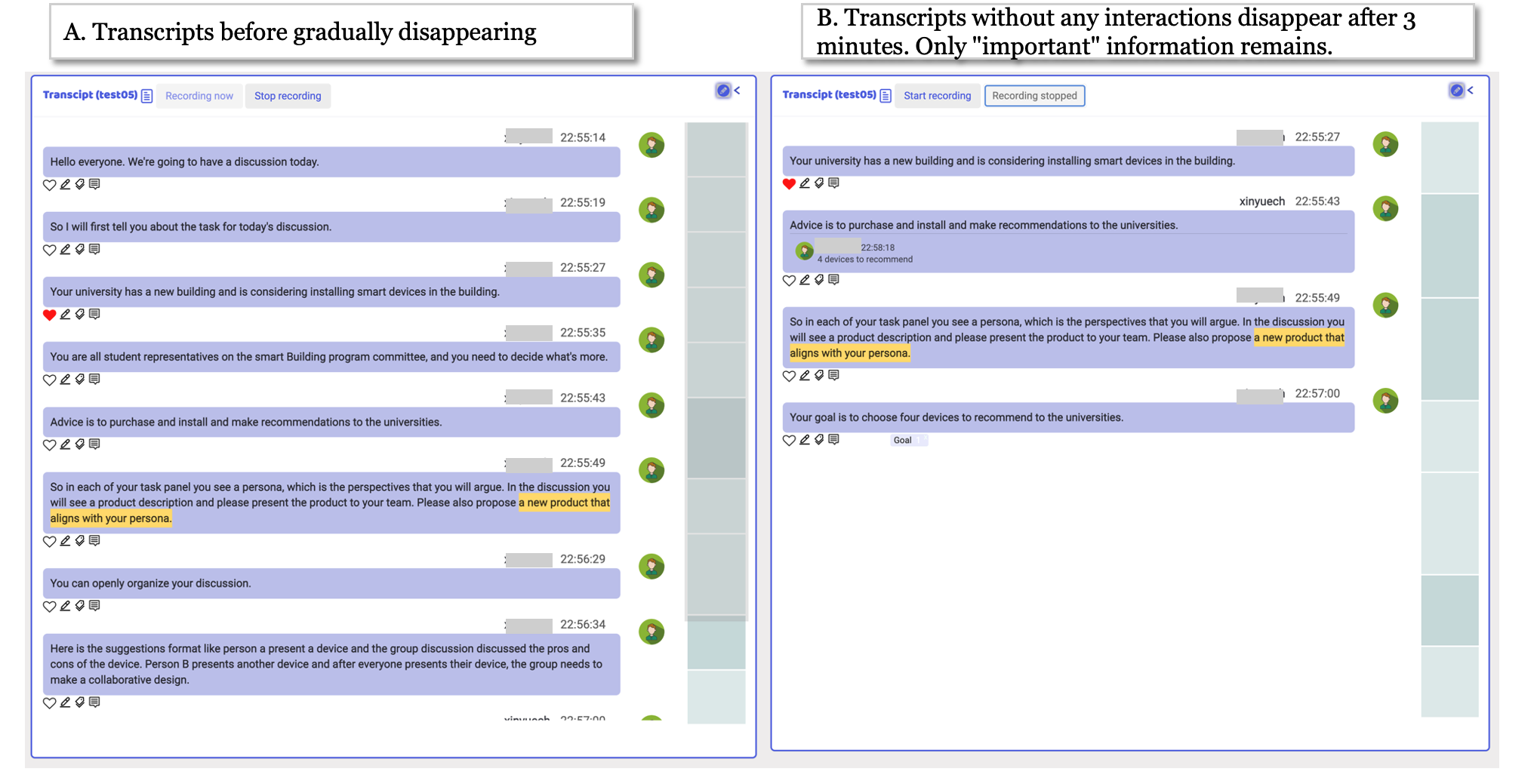} 
\caption{Transcript bubbles that do not have any user interactions will disappear after 3 minutes.}
\label{Fig.disappear}
\end{figure}

We take users' interaction as an indicator of the importance of a piece of text. For transcript bubbles that do not have any user interaction, they will disappear after 3 minutes, as shown in \ref{Fig.disappear}. The threshold of 3 minutes is decided through our pilot tests. 
Users in the pilot study agreed that this time threshold was enough for them to interact with the transcript bubbles. With the auto-disappearing feature, the content remaining on the screen is more concise, useful, and readable.

\subsubsection{Easy navigation of the transcript.}
To address the difficulty of searching information in transcripts as identified in the formative study, we designed an "Interaction Heatmap" Figure~\ref{Fig.system}(C4) and an "Interaction History Panel" Figure~\ref{Fig.system}(C5).
The "Interaction Heatmap" visualizes the density of interactions of the transcript in real time. 
The height of the grid in the Heatmap is computed proportionally to the height of the transcript bubble (which indicates the length of the message).
The color of the grid becomes deeper when there are more interactions on that transcript bubble. Any of the 5 interactions will deepen the color. Users can click on a grid to jump to a certain part of the conversation.
The "Interaction History Panel" allows users to index their own interactions. Specifically, users can filter content by the type of interactions, e.g., all my highlighted content. For all transcripts that have a tag, users can filter by the content of the tag, e.g., all the content that is tagged as "To-do". 
Both the Interaction Heatmap and the Interaction History Panel are designed to reduce the cognitive load required on the users when they review the transcript to look for information.

\subsection{System Implementation}
We implemented MeetScript using HTML, CSS, and Javascript in the front end and Django in the back end. We used Django Channels (the Django integration layer) and Django-Redis (Redis cache backend for Django) to handle the real-time updates in MeetScript. MeetScript uses the Jitsi Meet API \cite{jitsi_2021} to support videoconferencing (as shown in the video panel, Figure~ \ref{Fig.system}(A)). 
We used the Microsoft Azure SpeechSDK \cite{speech} to provide transcription because of its high accuracy (more than 90\%) \cite{xu2021benchmarking} and good community support for deployment and customization. MeetScript collects user audio from the client-side browser and then sends the transcript result to the database. We collect both interim transcripts and the finalized transcripts from the SpeechSDK. MeetScript updates the interim transcripts in real-time so that users see the transcription process while they are speaking. When the transcript is finalized, we will stabilize the transcript bubble and start a new one. When multiple people are speaking at the same time, the transcripts from different users will show in different bubbles.

\subsection{System Novelty}
In this section, we highlight the novelty of MeetScript. The idea of using live transcripts during video meetings is not new. Existing videoconferencing platforms such as Zoom and Google Meet have built-in live transcription. Additionally, one widely used transcription service, Otter.ai, provides note-taking functionalities where users can highlight texts or assign tasks on the transcript, as shown in Fig.\ref{Fig.otter}. However, both prior work and our formative study point out that such transcripts are unusable since they introduce extra cognitive load to users who are in a conversation. Existing transcription services are mainly designed to enhance meeting accessibility and to support post-meeting review. Challenges remain in leveraging live transcripts as a communication space for video meeting participants.

In the design of MeetScript, we aim to increase active participation through live transcripts and reduce the cognitive load required on users to process the information on the transcript. First, we visualize the discussion in a chat-alike interface, making it more readable and aligning with the verbal conversation flow. Second, the segmented transcript bubbles enable users to contextualize their interactions, unlike traditional parallel communication methods where chat messages are disconnected from the main discussion. Third, to further help users filter critical information and reduce cognitive load, we introduce an auto-disappearing mechanism based on user interaction, where transcript bubbles without interaction are auto-deleted over time. Lastly, we introduce two techniques, namely Interaction Heatmap and Interaction History Panel, to help users quickly navigate the conversation and locate their annotations. 
In summary, the design of MeetScript introduces techniques to make live transcripts interactive and reduce the cognitive load on users to process the information on the transcript. 

\section{Evaluation Study}
To evaluate the benefit of MeetScript in supporting active participation and understanding, we conducted a between-subjects experiment. The study is designed to answer the following research questions:
\begin{itemize}

    \item RQ1: Will MeetScript increase participation level in video meetings?
        \item[-] Hypothesis 1: MeetScript introduces additional participating channels without suppressing speech interactions.
    \item RQ2: How MeetScript users interact with the transcript to participate in video meetings?
    \item RQ3: Are transcript-based interactions perceived as valid and valuable contributions by team members?

        \item[-] Hypothesis 2: Participants in MeetScript perceive the transcript-based interactions as valid and valuable contributions.

    \item RQ4: Will MeetScript help people keep up with the conversation and enhance their understanding of the content being discussed?
        \item[-] Hypothesis 3: Groups negotiating a decision in MeetScript will show an increased understanding of the group decision and rationale.
        \item[-] Hypothesis 4:  Groups negotiating a decision in MeetScript will show an increased understanding of each individual member's perspectives and rationale. 
    \item RQ5: What are the perceived user experiences in MeetScript? What are the implications of developing video meeting support systems to enhance participation levels and mutual understanding?
\end{itemize}

\subsection{Study Procedure}
\begin{figure}[] 
\centering 
\includegraphics[width=\textwidth]{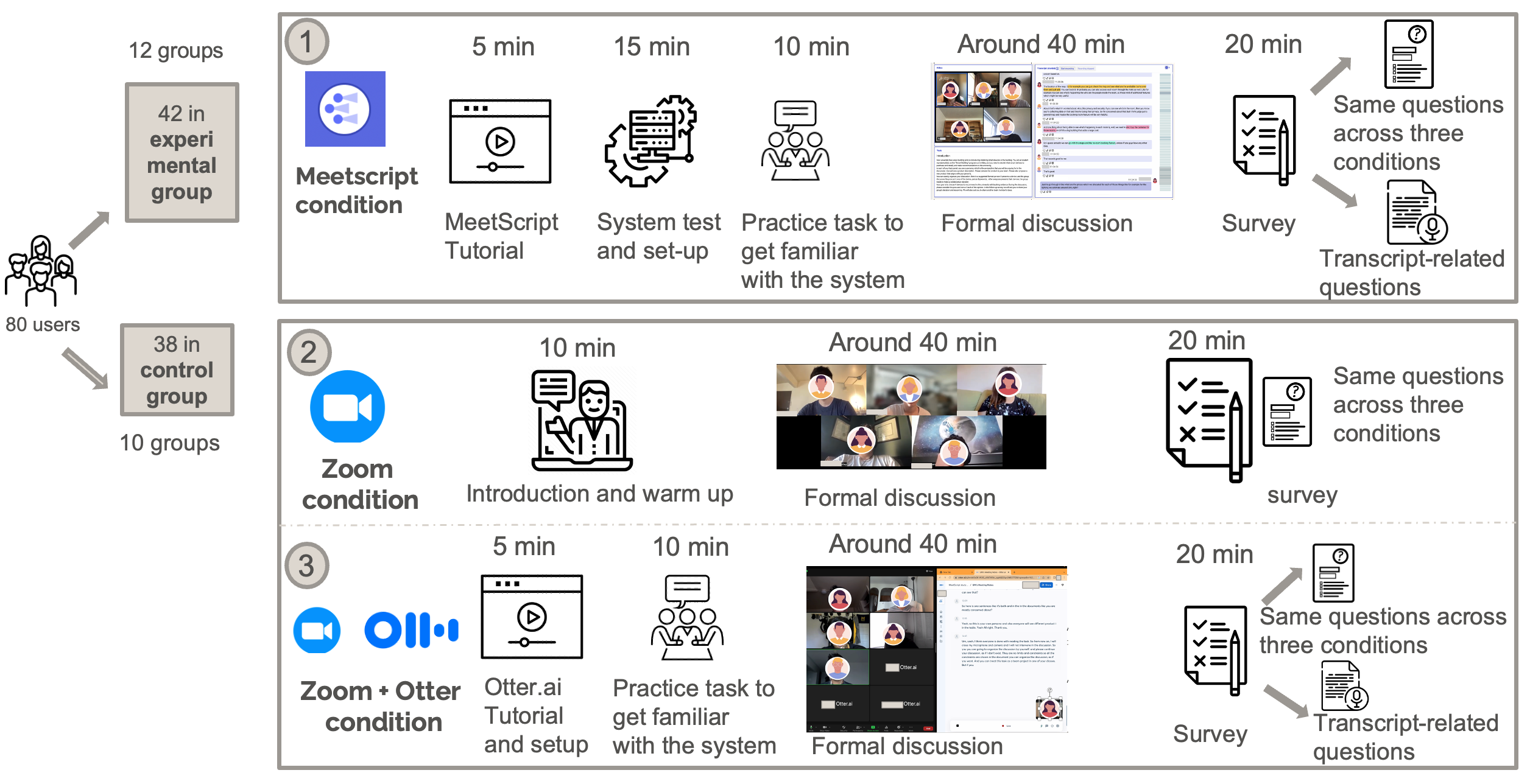} 
\caption{The participants were randomly assigned to the three conditions, MeetScript, Zoom, and Zoom+Otter.ai. All users across the three conditions had the same discussion task. Teams in MeetScript and Zoom+Otter.ai received a tutorial on how to use the system at the beginning. Then all teams had their discussion task. Post the discussion, and all teams answered a survey. The survey for MeetScript and Otter.ai conditions contained extra questions about the transcript.} 
\label{Fig.experiment}
\end{figure}

We conducted an IRB-approved between-subjects experiment. We compared MeetScript with two baseline conditions. In each condition, participants engaged in a decision-making task. We used system logs to quantify users' participation levels and administered a post-survey to collect both quantitative and qualitative data on users' understanding of the group negotiation process and their experiences using the system. The study procedure is shown in Fig~\ref{Fig.experiment}.

\subsubsection{Participants}
We recruited 88 participants (48 males, 40 females) through mailing lists at a large public university in the US. We selected participants who indicated that they were frequent users of Zoom for group meetings in the recruitment screening survey. Participants were randomly placed into groups of 4. Among the 22 groups, 12 were assigned to the MeetScript condition, and 10 were assigned to two baseline conditions. We introduced two baselines. First, a business-usual setup, where participants met over Zoom. They had the option of using any note-taking software, e.g., Google Docs. Second, we used Otter.ai to provide real-time transcription service to a Zoom meeting, where participants also had the chance to interact on the Otter.ai transcript panel. Otter.ai is a widely used transcription service with high accuracy\cite{adams2019artificial}.
Through the comparison of MeetScript with Zoom+Otter.ai, we can investigate whether the design of MeetScript can reduce the cognitive load on users to process the transcripts in real-time. We opted to conduct the evaluation study in small groups for two reasons. First, MeetScript is designed to foster active participation. In comparison to large-scale meetings, small groups necessitate more proactive engagement from each participant to ensure effective discussions \cite{bonito1997participation}.
Second, we explore the trade-offs between presenting users with more information without introducing unnecessary cognitive loads on them. Transcript-based interactions in large-scale meetings would bring in much more information since more people would contribute. Small group meetings offer a better opportunity for us to investigate the promises of transcript-based interactions to enhance active participation and mutual understanding.

Some participants did not show up to their assigned session, and we had a total of 80 participants (average age = 22, 46 male and 34 female). 42 participants were in the MeetScript condition (6 groups of 4 and 6 groups of 3), 18 participants were in the Zoom condition (3 groups of 4 and 2 groups of 3 participants), and 20 participants were in the Zoom+Otter.ai condition (5 groups of 4). Among all participants, 34 of them were undergrad students, 28 of them were master's students, and 18 of them were doctoral students.  

\subsubsection{Discussion Task}
To increase the level of information exchange, we followed task design under the hidden profile \cite{buder2008supporting, stasser1985pooling} and jigsaw scripting \cite{hinze2002jigsaw} paradigms. Each team member possesses unique pieces of information, and the group needs to negotiate to make a decision. 
The task resembles a course project discussion where a student team needs to brainstorm ideas and make a decision collaboratively. The goal of the task is to propose 4 smart-building devices to install in a university building. 
Each participant was assigned a persona and told they were most concerned about one perspective (cost, sustainability, ease of use, and data privacy). Each participant was also given an example smart-building device aligned with their persona. 
In the task, we asked the participants to present their assigned smart-building device to the team and brainstorm at least one more device that manifests their persona. With this setup, we expect to see at least 8 smart-building devices proposed in the discussion, and the team needs to decide on 4 devices to install. The team was asked to discuss freely, similar to the ways they had small group discussions before.

\subsubsection{Experimental Procedures}

In the MeetScript condition, participants were first given a 5-minute tutorial of MeetScript. Next, they were allotted about 15 minutes of system testing time to make sure everyone could use audio and video and have interactions on the transcript. This includes a 10-minute practice to try out the features of MeetScript. 
Similarly, in the Zoom+Otter.ai condition, participants were given a tutorial on the setup and a 10-minute practice period to explore the annotation features on Otter.ai. For the Zoom condition, since all the participants were familiar with Zoom, there was no need for a tutorial. The experimenter spent 5-10 minutes giving introductions.  We emphasized the importance of engaging in discussion with non-verbal interactions in all three conditions, where we showed participants the non-verbal interaction options through screen sharing during the onboarding time.

Participants across all three conditions then spent ~5 minutes reading the task. Afterward, they had ~40 minutes for free discussion. In the end, participants answered a survey. The survey is composed of three sections. The first section contains two open-ended questions asking participants to 1) summarize their group decision and rationale; 2) describe one specific team member's arguments. The second section contains Likert-style questions on participation and understanding levels of the group discussion. The third section contains transcript-specific questions. Only the participants in the MeetScript and the Zoom+Otter.ai conditions answered these questions.

The main difference between the three conditions lies in the time spent on introductions, tutorials, and onboarding of the system. The average length of the sessions is 90 minutes for MeetScript, 75.9 minutes for Zoom+Otter.ai, and 58.5 minutes for Zoom. In all conditions, before the discussion started, we asked participants to pay attention to their group decision outcome and rationale and individual team members' ideas since they'll be asked to describe them in the post-survey. All participants were paid at the rate of \$15/hr.

\subsection{Quantitative Outcome Measures}
\subsubsection{Measurement of Participation}
To address Hypothesis 1, two types of participation were counted: \textit{verbal participation} and \textit{non-verbal participation}. 
For all discussion sessions, we quantified users' verbal participation by the number of speech turns, the speaking time, and the number of words spoken during the meeting.  
To quantify non-verbal participation, we counted the number of chat messages, reactions, and polls used during the meeting. In addition, in MeetScript and Zoom+Otter.ai sessions, we counted the number of transcript-based interactions, including the number of highlights (both), comments (both), tags(MeetScript), likes (MeetScript), edits (MeetScript), and task assignment(Otter.ai).

\subsubsection{Measurement of Understanding}

To evaluate users' understanding of their discussion process, including the group decision and individual ideas (Hypothesis 3 and Hypothesis 4), we developed a coding manual to quantify participants' understanding as reflected in the post-survey. In the post-survey, participants were asked to answer two open-ended questions 1) summarize their group decision and rationale; 2) describe a specific team member's (e.g., Amy)'s arguments. The coding manual was used to assess the quality of answers to these two questions. 

One author first generated a "ground truth" answer for every session's final decision and their decision-making rationale by re-watching the discussion video. The "ground truth" answer was composed of a list of bullet points, including the products included in the final proposal and the rationale the team used to arrive at the final decision. Two authors coded all the answers and assigned each answer a score. 
For a team that included 4 products in the final proposal, we gave each product 1 point and the reasoning for choosing this product 1 point. The total possible points were 8. We awarded points if the user's answer aligned with the "correct" answer. We used the ratio of each participant's final score to the group's total score as a normalized understanding level. We applied the same coding process to participants' answers describing a specific team member's arguments. We provided examples of coded data in the appendix~\ref{coding}.

We want to highlight that the outcome measure used here indicated users' recollection of their decision-making process (with memory aids) more than in-situ understanding. 
In the MeetScript and Zoom+Otter.ai conditions, users could review the transcript when they completed the survey. In the Zoom condition, we asked the participants to use their business-as-usual video meeting setups. External note-taking tools are encouraged. Two groups used a Google Doc and referred to their notes when completing the survey. 

\subsection{Results}
\subsubsection{RQ1: Will MeetScript increase participation level?}

\textbf{MeetScript introduces additional participating channels without suppressing verbal interactions (Hypothesis 1 supported).}
Before checking whether MeetScript increases users' participation through the transcript, we need first to check whether MeetScript suppresses participants' verbal participation. We built a mixed-effect linear regression model, with the number of turns as the dependent variable and the conditions (MeetScript, Zoom, Zoom+Otter.ai) as the fixed effect. Considering teams' dynamics will influence users' individual performance, we included a random intercept for each team.
We found that the condition did not affect the number of verbal turns a participant had in a meeting (MeetScript versus Zoom+Otter.ai: coefficient = 1.202,  Std.Err = 5.382, p = 0.823 > 0.05, MeetScript versus Zoom: coefficient = -2.263, Std.Err = 5.439, p = 0.677 > 0.05 ). This suggests that MeetScript does not increase or decrease a participant's verbal participation in a video meeting. The descriptive statistics of users' verbal participation across the three conditions are shown in Table~\ref{table.statistics}.

\begin{table}[h]
\centering
\small
\begin{tabular}{lllllllll}
\toprule
 & \multicolumn{2}{l}{\textbf{MeetScript}} & \multicolumn{2}{l}{\textbf{Zoom}} & \multicolumn{2}{l}{\textbf{Zoom + Otter.ai}}  \\
                                   & mean    & std                  & mean    & std    & mean    & std       \\ \hline
verbal turns              & 22.2  & 10.5            & 21.1  & 9.7   & 24.3   & 7.8          \\
time spoken & 548.7 & 169.4               & 438.0 & 260.2    & 514.3 & 132.9     \\
words spoken           & 920.6 & 313.4             & 1031.6  & 603.1   & 1043.2  & 525.2   \\ 
transcript-based interaction & 10.83 & 16.87 & null & null & 0.42 & 0.83 \\
other non-verbal interaction (chat/reactions) & 0.28 & 0.53 & 0.55 & 0.92 & 0.63 & 0.95 \\
total non-verbal interaction & 11.11 & 16.82 & 0.55 & 0.92 & 1.0 & 1.49 \\
\bottomrule
\end{tabular}

\caption{\textbf{Descriptive statistics of the number of verbal and non-verbal interactions per person across three conditions.}}
\label{table.statistics}
\vspace{-2pc}
\end{table}

We then compared the number of non-verbal participants across the three conditions. For MeetScript and Zoom+Otter.ai teams, non-verbal participation was counted as the total number of transcript-based interactions and the number of chat messages and reactions sent during the meeting. For Zoom teams, non-verbal participation was the total count of chat messages, reactions, and polls sent during the meeting. Using a mixed-effect linear regression model, with the non-verbal participation counts as the dependent variable, the condition as the fixed effect, and a random intercept for each team, we observed that users in MeetScript had significantly more non-verbal interactions than those in Zoom + Otter.ai (coefficient = -9.54, Std.Err = 3.508, p = 0.007 < 0.05), and also significantly more non-verbal interactions than those in Zoom (coefficient = -10.291, Std.Err = 3.626, p = 0.005 < 0.05).

\textbf{Summary of findings in RQ1: The result shows that MeetScript increases users' non-verbal participation in the meeting through the transcript without suppressing their verbal participation.}

\subsubsection{RQ2: How MeetScript users use the transcript to participate in video meetings?}

\begin{figure}[h] 
\centering 
\includegraphics[width=\textwidth]{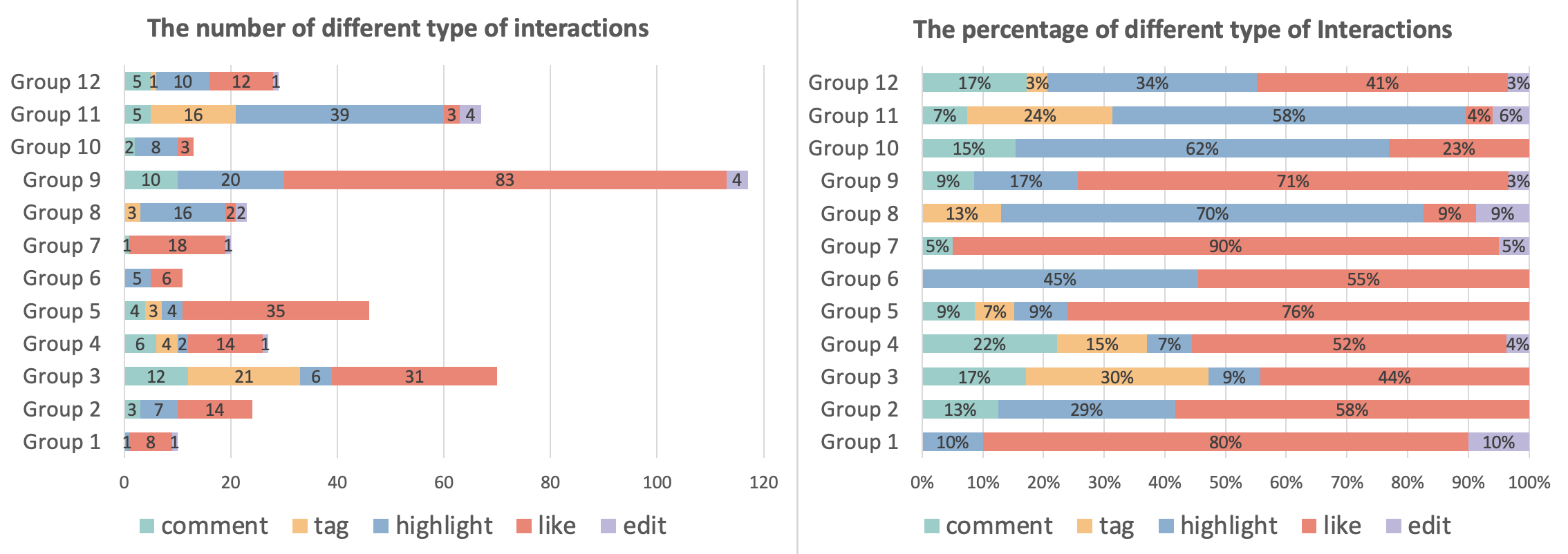} 
\caption{\textbf{Percentage of the usage of transcript-based interactions in MeetScript. }Interaction options with less effort to use (like and highlight) are more frequently used.}
\label{Fig.interaction}
\end{figure}

We further looked into how MeetScript users used the interaction options. Overall, most groups used the "like" feature heavily to keep track of good ideas. They also used the "highlight" feature to emphasize the key ideas. Other interactive options were used more infrequently, especially edits, possibly because users didn't have enough time to edit the transcript. The usage distribution is shown in Figure~\ref{Fig.interaction}. 

\begin{figure}[h] 
\centering 
\includegraphics[width=0.9\textwidth]{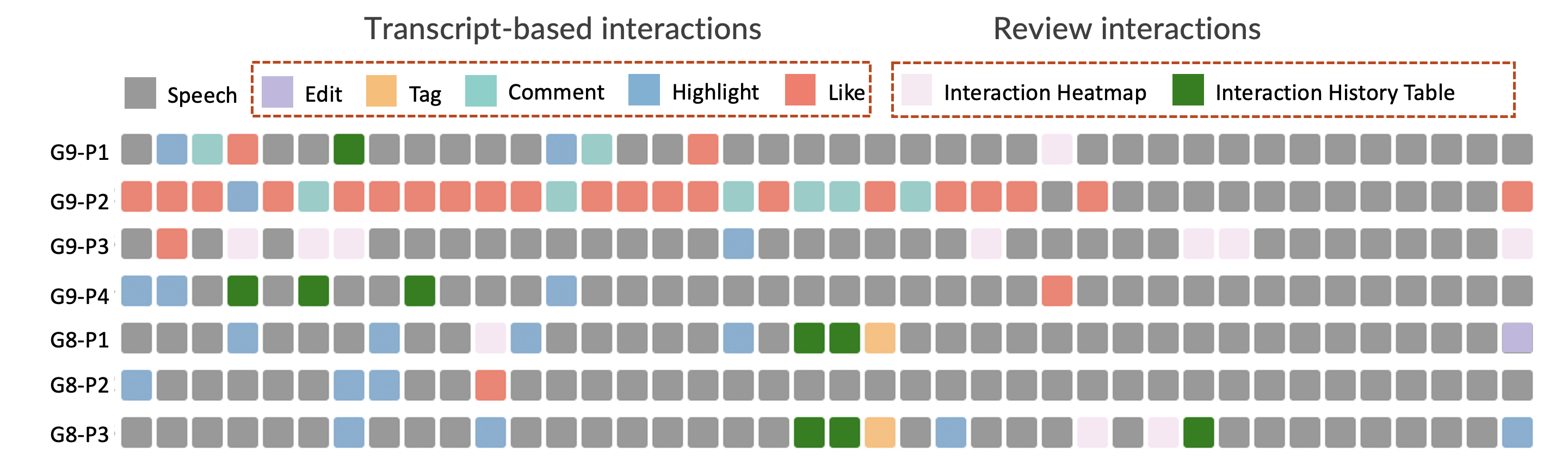} 
\caption{\textbf{The most frequently used interactions changed over time in two example sessions.} Participants used transcript-based interactions more often in the first three-quarters of the discussion when there were emerging new ideas.
}
\label{Fig.time}
\end{figure}

To understand how users used these interactions over the course of a meeting, we present two example sessions of MeetScript in Figure~ \ref{Fig.time}, which shows the interaction options that the attendants most frequently used over time. We split each discussion into 40 equal-duration slices and then computed the most frequently used feature in each slice. We combined log analysis with a video recording review to demonstrate what was happening in the group when the participants used MeetScript differently.

We found that in the first three-quarters of the meeting, the participants proposed more new ideas, and they used the transcript-based interactions more frequently during that time. P2 in Group 9 was very active in using "like" to express her attitudes towards others' ideas and using this to save important ideas. Everyone in these two groups used the highlight feature frequently. From the video recording, we observed that they often highlighted the pros and cons of a product that was proposed by others or highlighted details in their own transcript bubble that was ignored by others. 
We can also see that in Group 9, people started to comment more around the midpoint of the discussion. Through the video recording, we found that there was a participant who dominated the spoken channel, so others were commenting to add their opinions. 
Participants in Group 8 used the tag feature during the discussion to reach a consensus. Specifically, they assigned an "Agreed Product" tag to the products they decided to recommend. We also observed that participants used the Interaction Heatmap and the Interaction History Panel to help them navigate the transcript.
Specifically, G9-P3 used the Heatmap to navigate the transcript frequently during the meeting. Looking at their log data, we found that G9-P3 mainly scrolled back to the bubbles with more annotations. We also observed that two people in Group 8 frequently opened the Interaction History Panel to navigate to content that was highlighted by themselves. 

\textbf{Summary of findings in RQ2: MeetScript users primarily utilized the "like" and "highlight" interactions, which required lower effort. The transcript-based interactions were used more frequently when new ideas were proposed.}

\subsubsection{RQ3: Are transcript-based interactions perceived as valid and valuable contributions by team members?}
In previous sections, we addressed the utility of the transcript-based interactions provided by MeetScript to support participation. There arises the question of do meeting participants perceive their own or others' transcript-based interactions as a valuable contribution to the conversation.
Here we present people's ratings on their meeting experiences from the post-study survey. We performed the Kruskal-Wallis test to compare users' ratings between MeetScript and the two baseline conditions. We treated users' responses to a Likert-style question as ordinal variables. We found that people in the MeetScript condition rated the level of group participation to be significantly higher (p = 0.001 < 0.05) than those in Zoom (p = 0.002 < 0.05 in Dunn's posthoc test) and Zoom+Otter.ai conditions (p = 0.01 < 0.05 in Dunn's posthoc test). Participants in the Zoom and Zoom+ Otter.ai conditions showed similar perceived team participation levels (Q10, p = 0.48>0.05 in Dunn's posthoc test). This suggests that participants in MeetScript perceived the transcript-based interactions to be valid and valuable contributions by team members. (The descriptive statistics of the survey responses can be found in Table \ref{stat}).

In the post-survey, we designed questions that were specific to the usage of transcripts. Only users in the MeetScript and Zoom+Otter.ai conditions answered these questions. 
We ran the Mann-Whitney U test on users' ratings for the transcript-specific Likert-scale question for MeetScript and Zoom + Otter.ai conditions. MeetScript users agreed that they could use highlight/comments/tags to help them express their ideas, while most users didn't agree with the statement in the Zoom+Otter.ai condition (Q1, p=0.000 < 0.05). MeetScript users considered their team members were contributing through interactions on the transcript, whereas Otter.ai users did not think this way. (Q2, p=0.000 < 0.05). Users in MeetScript enjoyed using transcript-based interactions significantly more than those in Zoom+Otter.ai (Q3, p=0.000 < 0.05). 
User ratings to the survey questions are shown in Table~\ref{stat} (transcript-specific questions) \textbf{ (Hypothesis 2 supported)}. 

\textbf{Summary of the finding in RQ3: MeetScript users considered their team members' interactions on the transcript to be valuable contributions to the conversation. However, Zoom + Otter.ai users did not consider transcript-based interactions to be valuable contributions.}

\begin{table}[]
\small
\resizebox{\columnwidth}{!}{%
\begin{tabular}{p{3cm}p{1cm}p{5cm}lllllll}
\toprule
\multicolumn{3}{l}{\multirow{2}{*}{}} &
  \multicolumn{2}{c}{MeetScript} &
  \multicolumn{2}{c}{Zoom + Otter.ai} &
  \multicolumn{2}{c}{Zoom} &
   \\
\multicolumn{3}{l}{} &
  Mean &
  Std &
  Mean &
  Std &
  Mean &
  Std &
  p \\ \midrule
\multirow{9}{3cm}{\textbf{Transcript related questions: Scale 1-5 (1= strongly disagree, 5= strongly agree)}} 
 &
\textbf{Q1 }&
Using highlights/comments/task assignments helped me express my opinions. &
  3.88 &
  0.94 &
  2.67 &
  1.28 &
  \textbackslash{} &
  \textbackslash{} &
  0.000*** \\
 &
\textbf{Q2} &
When my teammates interacted with the transcript, it made me feel they were contributing to the conversation. &
  4.33 &
  0.65 &
  2.75 &
  1.16 &
  \textbackslash{} &
  \textbackslash{} &
  0.000*** \\
 &
\textbf{Q3} &
I enjoy using the non-verbal interaction ways provided by the tool to participate in the discussion. &
  3.97 &
  0.89 &
  2.8 &
  1 &
  \textbackslash{} &
  \textbackslash{} &
  0.000*** \\
 &
\textbf{Q4} &
When I was confused, the transcripts helped me understand my teammates' ideas. &
  4.19 &
  0.91 &
  3.12 &
  1.41 &
  \textbackslash{} &
  \textbackslash{} &
  0.004** \\
 &
\textbf{Q5} &
The transcripts increased the transparency of the conversation. &
  4.42 &
  0.73 &
  3.17 &
  1.48 &
  \textbackslash{} &
  \textbackslash{} &
  0.001** \\
 &
\textbf{Q6} &

The transcripts helped me understand my teammates' thoughts, e.g., what they think was important. &
  4.47 &
  0.59 &
  3.42 &
  1.31 &
  \textbackslash{} &
  \textbackslash{} &
  0.001** \\
 &
\textbf{Q7} &
The transcripts were distracting when I focused on talking with my teammates &
  2.31 &
  0.85 &
  3.12 &
  1.41 &
  \textbackslash{} &
  \textbackslash{} &
  0.099 \\
 &
\textbf{Q8} &
The annotations on the transcript helped me resolve misunderstandings and keep up with the conversation.&
  3.85 &
  0.97 &
  2.6 &
  1.04 &
  \textbackslash{} &
  \textbackslash{} &
  0.000*** \\
 &
\textbf{Q9} &
The transcripts helped me review previous discussions and synthesize the group's decision. &
  4.59 &
  0.49 &
  3.67 &
  1.34 &
  \textbackslash{} &
  \textbackslash{} &
  0.005** \\ \midrule
 
 \multirow{2}{3cm}{\textbf{General questions: Scale 1-10}}
 &
\textbf{Q10} &
  On a scale of 1-10, how would you rate the level of participation on your team? A score of 10 means everyone is perfectly focused and devoted to the conversation. &
  8.52 &
  1.08 &
  7.63 &
  0.95 &
  7.53 &
  2.06 &
  0.015* \\
 &
  \textbf{Q11} &
  On a scale of 1-10, how would you rate the level of mutual understanding achieved by your team? A score of   10 means you think everyone is perfectly on the same page. &
  8.65 &
  1.04 &
  7.42 &
  1.16 &
  7.47 &
  1.28 &
  0.001** \\ \bottomrule
\end{tabular}%
}
\caption{\textbf{The descriptive statistics on the survey responses across the three conditions.}}
\label{stat}

\end{table}

\label{question}

\subsubsection{RQ4: Will MeetScript help people keep up with the conversation and enhance their understanding of the content being discussed?}

\textbf{MeetScript Improves Users' Understanding of their Group Decision Reasoning (Hypothesis 3 partially supported).} 
We built a mixed-effects linear regression model with the participants' final score on their answer to the question summarizing the team's final decision and rationale as the dependent variable (The measurement can be found in 5.2.2), the condition (MeetScript, Zoom, and Zoom+Otter.ai) as the fixed effect, and a random intercept for each team. The results show a trend that users in MeetScript have a better understanding of the group decision (mean = 0.85, std = 0.17) than those in the Zoom+Otter.ai condition (mean = 0.75, std = 0.24), though the difference is not statistically significant (coefficient = -0.093,  p = 0.10). We observed a significantly higher level (coefficient = -0.167,  p = 0.005 < 0.05) of understanding in MeetScript than in the Zoom condition (mean = 0.68, std = 0.26).

\textbf{MeetScript Improves Users' Understanding of Individual Members' Ideas and Rationale (Hypothesis 4 supported).}
We built a mixed-effects linear regression model with the participants' final score on their answer to the question describing another team member's arguments and rationale as the dependent variable, the condition (MeetScript, Zoom, or Zoom+Otter.ai) as the fixed effect, and a random intercept for each team. 
We found that the condition had a significant effect on the level of understanding participants displayed. Specifically, MeetScript users' answers received higher score (mean = 0.84, std = 0.21) compared to those in Zoom (mean = 0.41, std = 0.29)( coefficient = -0.439,  p = 0.000 < 0.05), and those in Zoom+Otter.ai (mean = 0.61, std = 0.29) (coefficient = -0.225,  p = 0.004 < 0.05).
The two regression analyses show that users in MeetScript demonstrate a better recollection of their team's final decision and decision-making processes and of individual team members' arguments and rationale.

\textbf{MeetScript users found the transcript to be helpful for understanding the conversation}
We present people's ratings on survey questions about their understanding of the group discussion. We also performed Kruskal-Wallis tests to compare users' ratings between MeetScript and the two baseline conditions. 
We found that people in the MeetScript condition rated the level of mutual understanding achieved by their team to be significantly higher (Q11: p = 0.001 < 0.05) than those in the Zoom (p = 0.004 < 0.05 in Dunn's posthoc test) and the Zoom+Otter.ai (p = 0.010 < 0.05 in Dunn's posthoc test) conditions, while the understanding level between Zoom and Zoom+Otter.ai is similar (p = 0.082 > 0.05 in Dunn's posthoc test).

We then ran the Mann-Whitney U test to compare users' ratings on transcript-related understanding questions between the MeetScript and Zoom+Otter.ai conditions. MeetScript users agreed that the transcript helped them understand others' ideas, whereas users in the Zoom+Otter.ai condition did not agree with this statement. (Q6, p=0.001 < 0.05). 
Compared to Zoom+Otter.ai users, MeetScript users considered the transcript-based interactions to be more helpful in resolving misunderstandings (Q8, p=0.000 < 0.05). 
While MeetScript users considered the transcript to increase the transparency of the conversation, Zoom+Otter.ai did not think so (Q5, p=0.001 < 0.05). Apart from this, MeetScript users considered the transcripts helpful for them to review previous discussions and synthesize the group's decision, whereas Zoom+Otter.ai users' ratings were neutral (Q9, p=0.000 < 0.05).

\textbf{Summary of the findings in RQ4: MeetScript users exhibited an enhanced level of recollection of their group decision-making rationale and individual team member's arguments. They also found the transcript beneficial for understanding ideas, resolving misunderstandings, and reviewing discussions in contrast to users in the Zoom + Otter.ai condition.}

\subsubsection{RQ5: What are user experiences in MeetScript, and what are areas for improvement?}

To analyze the open-ended responses in the post-survey, we applied thematic analysis \cite{braun2006using}. Two authors did an initial round of open coding, discussed the initial codes, and then generated high-level themes. 

\textbf{MeetScript enables real-time, contextualized, and unobtrusive interactions.}
Most MeetScript participants shared that they found the transcript-based interactions helpful for active participation (35/42). First, they thought using features like comment/highlight/tag/save helped them interact with peers in a more flexible and natural way (13/42), as P13 said, \textit{"In Zoom to share any idea on someone else's idea or to add any comment, you need to wait for your turn or interrupt the person who is speaking, which at times is very uncomfortable and distracting. With this meeting tool, there is no need to disturb the speaker, and I can use the comment, save bubble, and highlight feature to share [my] thoughts."} Participants also felt that the transcript-based interaction enhanced the group atmosphere such that everyone was engaged in the discussion (5/42). As P9 said, \textit{"I like the ability to highlight, tag, and save messages, and I think that keeps users more engaged."} In contrast, users in the Zoom+Otter.ai condition used the transcript-based interactions less frequently. One of the reasons was that the annotations on the transcript were not well connected with the discussion context (5/20), as P62 said, \textit{"I don't think the annotation or comments in Otter.ai were very useful since it is not well linked to a section in the transcript, but it is on the side and separated as a takeaway list.''} Besides, most Otter.ai users mentioned that the speakers were not differentiated and the transcripts were disorganized, making it challenging to have extra interactions on the transcript (12/20); as P64 said, \textit{It did not identify who is who during the meeting. During the meeting, it could not identify the transcript based on the speakers. That made it difficult to follow who said what and make any annotations.}

Some participants in MeetScript thought commenting in the thread enabled them to add to the previous discussion without interrupting others (10/42). As P14 said, \textit{“I like the interaction functionality in MeetScript, which lets you leave the comment in any bubble that will not interrupt the person who is speaking.”} Additionally, some participants thought that comments based on the transcript made their participation more contextualized (8/42). As P4 said, \textit{"I use the comment to add to the previous discussion because the direct comment will not cause any confusion."} They found tagging to help them quickly summarize a conversation (6/42). As P5 said, \textit{"I like to use the tag as a quick annotation. And after we finish proposing all of the potential ideas, we use the tag to vote for which one we want to choose."} 
However, we observed overall negative sentiment towards interacting with the Otter.ai transcript during meetings (16/20), as P78 said, "It was just not very convenient to use Otter.ai when I was trying to both reference Zoom and argue for my own ideas. Otter.ai was not very useful in trying to annotate or comment on." Participants were also concerned about the visibility of their interactions through Otter.ai (4/20), as P75 said, \textit{"The transcript moves along and continues scrolling even when I have text highlighted and am working on a comment, which made annotating things being said recently difficult."}

\textbf{MeetScript helps users keep up with the conversation in real-time}
Participants in MeetScript indicated that they were less afraid of missing important information during the meeting (18/42). P18 said, \textit{"MeetScript allows you to catch up with what you missed, that you won't have the whole group waiting for you to understand their ideas."} Participants thought the transcripts and annotations in MeetScript helped them understand others' priorities (10/42). P39 said, \textit{ "I like the interactive features. They not only help me to organize the conversation but also help me to know what other people are doing/thinking."} Users found the highlighting feature to be useful in understanding others' ideas (7/42). As P36 said, \textit{"The tools included in MeetScript, especially the transcript and highlighting around the transcript, are very useful! It definitely cleared things up when I misheard someone or if I didn't get to the key point."} 
Similarly, some users in Otter.ai also found the transcripts helpful for keeping track of the conversation (8/20), but many of them felt reading the transcript in Otter.ai to be distracting (15/20), as P67 said, \textit{"Zoom \& otter.ai is distracting to use together because there is a lot going on. Although it only took me a couple of extra seconds to use these features, these few seconds took my attention away from the conversation, and therefore, I lost track of it.''}

\textbf{MeetScript benefits post-meeting review and reflection}
MeetScript users found the transcript helpful for them to review previous discussions and synthesize the group decision after a meeting (30/42).
Most users thought the transcript in MeetScript to be super accurate and performed better than other meeting tools they had used (24/42). As P32 said, \textit{"The transcription was outstandingly accurate. I did not notice that [any] of my words were wrong, even though I have an accent." } The accuracy of the transcript influenced how users received the transcript in post-meeting reviews, and many users in the Zoom+Otter.ai condition complained about the errors in the transcript (10/20), as P74 said, \textit{"A lot of the words were very wrong, to the point where I could not comprehend what was being said.''}
Participants in Meetscript felt being able to review the annotated transcript after a meeting helped them keep track of what had happened (22/42). 
As P32 said, `\textit{`I love being able to go back and review important info without having to have a dedicated note taker in the meeting. "} MeetScript users found it easy to navigate an annotated transcript and find important information (6/42). 
P24 mentioned,\textit{ "The transcript, transcript editing, and mini-map make reviewing previous conversations much easier, and they are not available anywhere else, which makes MeetScript the go-to choice for every scenario."} On the other hand, most participants in Zoom+Otter.ai cannot recognize the benefits of reviewing transcripts post a meeting, with one of the reasons being a lack of navigation support (5/20), as P69 mentioned, \textit{"I couldn't see who said what in the transcripts which made it hard to look back at the conversations. Besides, Otter.ai seemed a bit hard to navigate and find certain information in such a long document. "}

\textbf{Users are interested in using MeetScript in academic and workplace contexts.}
In the survey, we asked the users to compare MeetScript with other video meeting setups they had used (e.g., Zoom) to see what they liked or disliked about MeetScript, and in what scenarios they would prefer to use MeetScript. Among the 42 MeetScript users, 4 users said they preferred to use Zoom for all future meetings since they were used to it and it was more stable than MeetScript. The other 38 users said they'd prefer to use MeetScript in some contexts. Among them, 10 users explicitly said that they preferred to use MeetScript in all meeting scenarios if existing UI bugs were resolved, and the transcript became more accurate. 
Some participants thought MeetScript would be more useful for online lectures and presentations in academic contexts. In these circumstances, it is not convenient for everyone to speak up when they want. P15 said \textit{"I would prefer Zoom for more active discussions and MeetScript for more presentation/lecture types where it is not possible for participants to speak during the presentation yet they still want to share their ideas/highlight.".} Some users think MeetScript would be good for project-based group discussions. 
They commented that saving the content is especially important for their project meetings. As P40 said, \textit{"I would prefer this for group meetings as this would allow me to keep track of all the ideas, especially for the report. ".} Participants also see the potential of using MeetScript in cross-cultural group discussion as P32 said, \textit{"I would prefer this system in foreign language settings, or when talking with people with strong accents."} Some participants mentioned the ability to use this tool to do qualitative research and interviews, as P33 said, \textit{"This system would greatly benefit qualitative researchers or journalists who conduct interviews, etc. I liked the tag feature a lot; the idea of it seems very beneficial, especially in the research setting."} On the other hand, participants expressed preferring Zoom for casual chats when there is no need to keep track of the conversation. In contrast, most participants in the Zoom+ Otter.ai sessions said they would prefer to use Zoom only since the transcript was more distracting rather than helpful.

\textbf{Challenges in Using MeetScript.}
While users reported many benefits of using MeetScript, they also mentioned challenges they faced. First, a few participants (4/42) felt distracted when interacting with the transcript during the discussion. As P18 mentioned, \textit{"I think it was a little difficult to monitor and interact with the transcripts while also being an active listener and participator."} While in Zoom+Otter.ai sessions, most participants (15/20) mentioned the transcripts to be distracting. This suggests the mechanisms we designed to reduce the cognitive load and information density in the transcript were helpful, but there are still opportunities to improve it for more natural and richer interaction.
Some participants mentioned there was a learning curve in using the system. 
P10 said \textit{"I was really confused with how it worked at first, so the learning curve was a bit steep. But then I feel I like the saving and highlighting features, and it allows more interaction than other platforms that I have used. However, because there are so many new features for MeetScript, it can be overwhelming, especially for a new user. "} 
Second, a few users disliked how the transcript bubbles without interactions disappeared.
P9 said, \textit{"I think the transcribed messages should not disappear after 3 minutes, though, because if someone forgot to interact with a message, the message would disappear. It can be better if it is folded or unfolded rather than completely disappear."} In future work, we plan to provide more user control so that users can fold or unfold transcript bubbles when needed.
Third, some users mentioned they were uncertain whether other team members noticed their annotations on the transcript. P7 said \textit{"When I add a comment or highlight, I don't think people really noticed what I said or highlighted. "} In future work, we plan to experiment with notification mechanisms to make the transcript-based interactions more explicit.

\textbf{Design Suggestions for MeetScript}
Users also provided valuable feedback for us to continue improving the MeetScript system. First, for easier navigation, participants wanted to have additional ways to filter information. P25 said, \textit{"Each person's transcript should appear in a different color, and a filter should be added so that I can find a specific person's words more easily."} They also hope to see more information on the interaction heat map to help them better locate information on the transcript. P33 said, \textit{"I think it would also be really nice to be able to see things on the minimap, like the tags, or a symbol that shows comments on one square, etc."} 
Some users expected to see a more concise transcript. P26 mentioned, \textit{ "
I think one way to further improve this app is to shorten the transcript and just keep the useful information. Many words are not needed for understanding."} Meanwhile, participants proposed additional interaction options they'd like to use, for example, emojis. P2 said, `\textit{`I think it would be nice if there was a feature that allowed peers to agree with each other on the transcripts as opposed to having to comment, so something like a thumbs-up button, and this would also be able to keep track of the popularity of ideas."} 

\textbf{Summary of findings in RQ5: MeetScript was found to enhance real-time, contextualized, and unobtrusive interactions, improve understanding of group discussions, and facilitate post-meeting review, in contrast to real-time transcripts provided by traditional videoconferencing platforms. Users have suggested enhancing the MeetScript system by implementing additional navigation and filtering options, condensing transcripts to retain only essential information, and adding interactive features like emojis and thumbs-up buttons for agreement and tracking the popularity of ideas. Users also mentioned challenges, including the initial learning curve in using the system and distractions.}

\section{Discussion}

\subsection{Extending the Communication Space in Video Meetings to Support Active Participation and Understanding}

Our evaluation study provides quantitative and qualitative evidence that MeetScript provides effective parallel participation channels in group video meetings. By comparing two baseline conditions (Zoom and Zoom+Otter.ai), we show that meeting attendants have statistically significant more non-verbal participation when using MeetScript. MeetScript users generally considered the system to be usable, and across the experimental sessions, participants came up with a variety of creative ways to use the transcript-based interactions. The interactions that require lower effort, specifically "highlight" and "like" are more heavily used than other interaction options, while more elaborate annotations or comments can provide valuable insights and promote deeper discussions. Importantly, MeetScript users regard their team members' transcript-based interactions as valuable contributions to the discussion.  These findings align with the study on how student collaborative annotation can enhance collaboration and learning outcomes in education settings \cite{fang2021notecostruct}, and we extend these studies into group video meeting scenarios. 

In addition, we see promising results that MeetScript improves meeting participants' recollection of their team's decision-making processes and individual team members' ideas compared to the two baselines. 
Compared to a widely adopted transcription service Otter.ai, MeetScript users display entirely different sentiments toward using transcripts to facilitate video meetings. MeetScript users find the transcripts to be helpful in increasing meeting transparency, resolving misunderstandings, understanding their teammates' ideas, and reviewing and synthesizing prior discussions. However, Otter.ai users mostly disagreed with such statements.   

We consider MeetScript to be a further step in investigating how to extend the communication space in video meetings, which resonates with other work on extending the parallel communication channel alongside video conferences \cite{sarkar2021promise, yang2022catchlive, warner2023slidespecs}. We show evidence that live transcripts can be used as an extra communication space when the system visualizes transcripts through a readable interface that aligns with the conversation flow, invites contextualized and low-effort user interactions, and embeds information filtering mechanisms that help users locate important information.

\subsection{Design Implications on Using Transcripts to Facilitate Video Meetings.} The design and development of MeetScript suggest that when we provide a real-time transcript to users, which is added information, it is critical to design techniques that keep the essential information and helps users better navigate the added information. The goal is to provide an appropriate amount of information while not overwhelming the users.
Specifically, the three techniques to reduce information density we implemented in MeetScript were well-received by participants to help them reduce the cognitive load in processing the transcript. Based on the evaluation study findings, we propose four design implications for researchers and practitioners when they develop video meeting platforms facilitated by real-time transcripts or collaborative sense-making tools. 

First, it is important to address the issue of distraction faced by some participants when interacting with a transcript (or any external documents) during a video meeting. In MeetScript, we found visualizing the transcript document that resembles the discussion process to be a useful strategy \cite{zhang2018co}. Future work may explore the integration of AI techniques to enhance the functionality of MeetScript, e.g., summarization-based techniques \cite{hong2014repository, lins2019cnn, yang2022catchlive} to extract important ideas for users. 
Specifically, user annotations can serve as an important input for summarization algorithms to produce more reliable outcomes \cite{yang2022catchlive}. For example, prior work has developed human-AI techniques to generate meeting summaries\cite{cheng2022mapping} and employed 
user-generated tags and pre-defined hashtags to structure long conversations\cite{zhang2018making, geller2020confused}. 

Second, in the evaluation study, we found that some users preferred to receive a condensed version of the transcript, whereas others wanted the flexibility to go back to the content that was filtered out by the system. We recommend giving users more control over what information they read and retrieve in future system design. For example, the transcript panel may only show a skeleton of information, while the users can hover over it to get more details. 

Third, it is exciting that many MeetScript users considered transcript-based interactions to be valuable contributions to the conversation and helped them keep up with the discussion. Moreover, in the post-survey, users expressed the desire for their interactions on the transcript to be noticed and received by the team. Future design could consider how to make transcript-based interactions more visible, e.g., through explicit notification, as a way to encourage diverse participation channels for online video meetings. 

Lastly, users also gave feedback to further improve the usability of the system. For example, introducing advanced navigation and filtering techniques. Users also wanted to have additional interaction options and use emojis to express agreement or support.

\subsection{Adapt MeetScript to Support Diverse Usage Scenarios}
Although our need-finding and evaluation studies are mainly focused on video meetings with relatively small group sizes (3-5), we envision MeetScript can be extended to more diverse scenarios. 
Based on user feedback, we found the great potential of using MeetScript in educational settings, especially in lectures. Some problems the participants mentioned, e.g., people cannot keep up with long monologues, it is hard to receive information from speakers passively, and it feels intimidating to interrupt, are similar to students' challenges to keep up with lectures. While in educational settings, there is a stronger need for students to understand the content and learn from each other \cite{wang2017contrasting, adamson2014towards}. 
Previous studies in educational contexts have found that activities such as creating concept maps for MOOCs videos benefit student learning more than passively receiving a lecture \cite{chi2014icap}. In future work, we hope to extend MeetScript as a collaborative annotation tool for lectures and video tutorials to help students actively learn during and post lectures.

Previous studies emphasized the need to support non-native speakers to keep up with rapid conversations in a new language \cite{duan2021bridging}. Some researchers have explored the usage of clarification agents to help non-native speakers to ask questions \cite{duan2018let}. Others have explored incorporating automated transcripts to help non-native speakers to solve confusion with native speakers' annotation \cite{gao2014effects}. MeetScript may provide more channels for non-native speakers to participate and make contextualized annotations. 

We also envision MeetScript to support people with hard of hearing in meetings. Deaf and Hard-of-Hearing (DHH) users face accessibility challenges during in-person and remote meetings \cite{seita2021deaf}. Previous studies found that adding highlights on keywords in lecture videos could improve deaf or hard-of-hearing students' learning experiences \cite{kafle2021deaf}. Previous research also mentioned the need for improving the quality of automatically-generated transcripts to help DHH users in meetings \cite{seita2021deaf}. We envision non-DHH users may help edit and correct transcripts through systems such as MeetScript to help DHH users.

\subsection{Limitation}
First, the presented study of MeetScript focused on new teams where team members did not know each other before. This resembles the experience for many in-class project discussions and ad-hoc project meetings in workplaces \cite{stewart2020beyond}. The results found in this work would apply to new and non-established teams. We will leave it to future work to explore how such systems influence the collaboration dynamics in established teams. And it would be valuable to investigate the long-term effects of MeetScript on team collaboration and decision-making, as well as the potential impact on remote work practices. 
Second, although we found positive results that teams using MeetScript had better recollections of their teams' decision rationale and individual team members' ideas, recollection does not equal in-situ understanding since users had a reference panel when they answered the survey questions. In future work, we would like to further evaluate whether MeetScript can enhance people's in-situ understanding of the content and lead to higher-quality collaborative outcomes.  
Third, we recruited the participants of the evaluation study from a mailing list at one university, and we ran the evaluation study with small group meetings. In future evaluation and deployment studies, we hope to recruit a more diverse group of participants. Further research could also examine the scalability of MeetScript in larger group settings and across different types of organizations, industries, and use cases.   
Fourth, the sample size among the three conditions was unequal since it is challenging to find 12 groups of 4 for all three conditions. We think having an equal number of participants among the three conditions will add to the robustness and power of the statistics. Last, the current prototype system had remaining usability issues, e.g., delays or errors in live-transcript at times. There is also a big learning curve for people to learn to use and adapt to this system. These factors may have influenced people's usage and interactions with the system. In the next step, we plan to run a longitudinal study to study people's behaviors once they become familiar with the system.

\section{Conclusion}
We present MeetScript, a system that offers parallel participation channels for online meeting participants through a real-time interactive meeting transcript. MeetScript visualizes a conversation through a chat-alike interface, enables users to make annotations contextualized in the discussion, and introduces mechanisms to help users filter important information and easily navigate the transcript. We investigate methods to present live transcripts that do not add extra cognitive load on users while they are having a conversation. 
In an evaluation with 80 users, we demonstrate that MeetScript significantly increases the non-verbal participation of video meeting users. Additionally, MeetScripts users show better recollection of their teams' decisions making processes and individual team members' arguments. Participants find the transcripts in MeetScript to be more useful and understandable compared to a baseline condition using Otter.ai. We discuss the promises of using live transcripts as an extra communication space for meeting attendants and the implications of supporting active participation and understanding while not overwhelming users with extra information.

\bibliographystyle{ACM-Reference-Format}
\bibliography{references.bib}


\begin{thebibliography}{97}


\ifx \showCODEN    \undefined \def \showCODEN     #1{\unskip}     \fi
\ifx \showDOI      \undefined \def \showDOI       #1{#1}\fi
\ifx \showISBNx    \undefined \def \showISBNx     #1{\unskip}     \fi
\ifx \showISBNxiii \undefined \def \showISBNxiii  #1{\unskip}     \fi
\ifx \showISSN     \undefined \def \showISSN      #1{\unskip}     \fi
\ifx \showLCCN     \undefined \def \showLCCN      #1{\unskip}     \fi
\ifx \shownote     \undefined \def \shownote      #1{#1}          \fi
\ifx \showarticletitle \undefined \def \showarticletitle #1{#1}   \fi
\ifx \showURL      \undefined \def \showURL       {\relax}        \fi
\providecommand\bibfield[2]{#2}
\providecommand\bibinfo[2]{#2}
\providecommand\natexlab[1]{#1}
\providecommand\showeprint[2][]{arXiv:#2}

\bibitem[\protect\citeauthoryear{??}{ott}{[n.d.]}]%
        {otter}
 \bibinfo{year}{[n.d.]}\natexlab{}.
\newblock \bibinfo{title}{Otter {Voice} {Meeting} {Notes}}.
\newblock
\newblock
\urldef\tempurl%
\url{https://otter.ai}
\showURL{%
\tempurl}


\bibitem[\protect\citeauthoryear{??}{spe}{[n.d.]}]%
        {speech}
 \bibinfo{year}{[n.d.]}\natexlab{}.
\newblock \bibinfo{title}{Speech to text – audio to text translation:
  Microsoft azure}.
\newblock
\newblock
\urldef\tempurl%
\url{https://azure.microsoft.com/en-us/services/cognitive-services/speech-to-text/}
\showURL{%
\tempurl}


\bibitem[\protect\citeauthoryear{??}{jit}{2021}]%
        {jitsi_2021}
 \bibinfo{year}{2021}\natexlab{}.
\newblock \bibinfo{title}{Jitsi API: How to ADD video meetings to your site}.
\newblock
\newblock
\urldef\tempurl%
\url{https://jitsi.org/api/}
\showURL{%
\tempurl}


\bibitem[\protect\citeauthoryear{Adams}{Adams}{2019}]%
        {adams2019artificial}
\bibfield{author}{\bibinfo{person}{Nathan-Ross Adams}.}
  \bibinfo{year}{2019}\natexlab{}.
\newblock \showarticletitle{How artificial intelligence< currently> works}.
\newblock \bibinfo{journal}{\emph{Becoming AI}} (\bibinfo{year}{2019}).
\newblock


\bibitem[\protect\citeauthoryear{Adamson, Dyke, Jang, and Ros{\'e}}{Adamson
  et~al\mbox{.}}{2014}]%
        {adamson2014towards}
\bibfield{author}{\bibinfo{person}{David Adamson}, \bibinfo{person}{Gregory
  Dyke}, \bibinfo{person}{Hyeju Jang}, {and} \bibinfo{person}{Carolyn~Penstein
  Ros{\'e}}.} \bibinfo{year}{2014}\natexlab{}.
\newblock \showarticletitle{Towards an agile approach to adapting dynamic
  collaboration support to student needs}.
\newblock \bibinfo{journal}{\emph{International Journal of Artificial
  Intelligence in Education}} \bibinfo{volume}{24}, \bibinfo{number}{1}
  (\bibinfo{year}{2014}), \bibinfo{pages}{92--124}.
\newblock


\bibitem[\protect\citeauthoryear{Aseniero, Constantinides, Joglekar, Zhou, and
  Quercia}{Aseniero et~al\mbox{.}}{2020}]%
        {aseniero2020meetcues}
\bibfield{author}{\bibinfo{person}{Bon~Adriel Aseniero},
  \bibinfo{person}{Marios Constantinides}, \bibinfo{person}{Sagar Joglekar},
  \bibinfo{person}{Ke Zhou}, {and} \bibinfo{person}{Daniele Quercia}.}
  \bibinfo{year}{2020}\natexlab{}.
\newblock \showarticletitle{MeetCues: Supporting Online Meetings Experience}.
\newblock \bibinfo{journal}{\emph{arXiv preprint arXiv:2010.06259}}
  (\bibinfo{year}{2020}).
\newblock


\bibitem[\protect\citeauthoryear{{Backlinko}}{{Backlinko}}{2021}]%
        {Zoomusers}
\bibfield{author}{\bibinfo{person}{{Backlinko}}.}
  \bibinfo{year}{2021}\natexlab{}.
\newblock \bibinfo{title}{Zoom User Stats}.
\newblock
\newblock
\urldef\tempurl%
\url{https://backlinko.com/zoom-users}
\showURL{%
\tempurl}


\bibitem[\protect\citeauthoryear{Bailey and Konstan}{Bailey and
  Konstan}{2006}]%
        {bailey2006need}
\bibfield{author}{\bibinfo{person}{Brian~P Bailey} {and}
  \bibinfo{person}{Joseph~A Konstan}.} \bibinfo{year}{2006}\natexlab{}.
\newblock \showarticletitle{On the need for attention-aware systems: Measuring
  effects of interruption on task performance, error rate, and affective
  state}.
\newblock \bibinfo{journal}{\emph{Computers in human behavior}}
  \bibinfo{volume}{22}, \bibinfo{number}{4} (\bibinfo{year}{2006}),
  \bibinfo{pages}{685--708}.
\newblock


\bibitem[\protect\citeauthoryear{Beers, Boshuizen, Kirschner, and
  Gijselaers}{Beers et~al\mbox{.}}{2006}]%
        {beers2006common}
\bibfield{author}{\bibinfo{person}{Pieter~J Beers}, \bibinfo{person}{Henny~PA
  Boshuizen}, \bibinfo{person}{Paul~A Kirschner}, {and} \bibinfo{person}{Wim~H
  Gijselaers}.} \bibinfo{year}{2006}\natexlab{}.
\newblock \showarticletitle{Common ground, complex problems and decision
  making}.
\newblock \bibinfo{journal}{\emph{Group decision and negotiation}}
  \bibinfo{volume}{15}, \bibinfo{number}{6} (\bibinfo{year}{2006}),
  \bibinfo{pages}{529--556}.
\newblock


\bibitem[\protect\citeauthoryear{Berry}{Berry}{2019}]%
        {berry2019role}
\bibfield{author}{\bibinfo{person}{Sharla Berry}.}
  \bibinfo{year}{2019}\natexlab{}.
\newblock \showarticletitle{The role of video and text chat in a virtual
  classroom: How technology impacts community}.
\newblock In \bibinfo{booktitle}{\emph{Educational technology and resources for
  synchronous learning in higher education}}. \bibinfo{publisher}{IGI Global},
  \bibinfo{pages}{173--187}.
\newblock


\bibitem[\protect\citeauthoryear{bin Rosawi}{bin Rosawi}{2020}]%
        {bin2020Zoom}
\bibfield{author}{\bibinfo{person}{Khairul~Anwar bin Rosawi}.}
  \bibinfo{year}{2020}\natexlab{}.
\newblock \showarticletitle{Zoom user guide}.
\newblock \bibinfo{journal}{\emph{International Journal of TESOL Studies}}
  \bibinfo{volume}{2}, \bibinfo{number}{2} (\bibinfo{year}{2020}),
  \bibinfo{pages}{174--183}.
\newblock


\bibitem[\protect\citeauthoryear{Bonito and Hollingshead}{Bonito and
  Hollingshead}{1997}]%
        {bonito1997participation}
\bibfield{author}{\bibinfo{person}{Joseph~A Bonito} {and}
  \bibinfo{person}{Andrea~B Hollingshead}.} \bibinfo{year}{1997}\natexlab{}.
\newblock \showarticletitle{Participation in small groups}.
\newblock \bibinfo{journal}{\emph{Annals of the International Communication
  Association}} \bibinfo{volume}{20}, \bibinfo{number}{1}
  (\bibinfo{year}{1997}), \bibinfo{pages}{227--261}.
\newblock


\bibitem[\protect\citeauthoryear{Braun and Clarke}{Braun and Clarke}{2006}]%
        {braun2006using}
\bibfield{author}{\bibinfo{person}{Virginia Braun} {and}
  \bibinfo{person}{Victoria Clarke}.} \bibinfo{year}{2006}\natexlab{}.
\newblock \showarticletitle{Using thematic analysis in psychology}.
\newblock \bibinfo{journal}{\emph{Qualitative research in psychology}}
  \bibinfo{volume}{3}, \bibinfo{number}{2} (\bibinfo{year}{2006}),
  \bibinfo{pages}{77--101}.
\newblock


\bibitem[\protect\citeauthoryear{Buder and Bodemer}{Buder and Bodemer}{2008}]%
        {buder2008supporting}
\bibfield{author}{\bibinfo{person}{J{\"u}rgen Buder} {and}
  \bibinfo{person}{Daniel Bodemer}.} \bibinfo{year}{2008}\natexlab{}.
\newblock \showarticletitle{Supporting controversial CSCL discussions with
  augmented group awareness tools}.
\newblock \bibinfo{journal}{\emph{International Journal of Computer-Supported
  Collaborative Learning}} \bibinfo{volume}{3}, \bibinfo{number}{2}
  (\bibinfo{year}{2008}), \bibinfo{pages}{123--139}.
\newblock


\bibitem[\protect\citeauthoryear{Cao, Lee, Iqbal, Czerwinski, Wong, Rintel,
  Hecht, Teevan, and Yang}{Cao et~al\mbox{.}}{2021}]%
        {cao2021large}
\bibfield{author}{\bibinfo{person}{Hancheng Cao}, \bibinfo{person}{Chia-Jung
  Lee}, \bibinfo{person}{Shamsi Iqbal}, \bibinfo{person}{Mary Czerwinski},
  \bibinfo{person}{Priscilla Wong}, \bibinfo{person}{Sean Rintel},
  \bibinfo{person}{Brent Hecht}, \bibinfo{person}{Jaime Teevan}, {and}
  \bibinfo{person}{Longqi Yang}.} \bibinfo{year}{2021}\natexlab{}.
\newblock \bibinfo{title}{Large Scale Analysis of Multitasking Behavior During
  Remote Meetings}.
\newblock
\newblock
\showeprint[arxiv]{2101.11865}~[cs.CY]


\bibitem[\protect\citeauthoryear{Chandrasegaran, Bryan, Shidara, Chuang, and
  Ma}{Chandrasegaran et~al\mbox{.}}{2019}]%
        {chandrasegaran2019talktraces}
\bibfield{author}{\bibinfo{person}{Senthil Chandrasegaran},
  \bibinfo{person}{Chris Bryan}, \bibinfo{person}{Hidekazu Shidara},
  \bibinfo{person}{Tung-Yen Chuang}, {and} \bibinfo{person}{Kwan-Liu Ma}.}
  \bibinfo{year}{2019}\natexlab{}.
\newblock \showarticletitle{TalkTraces: real-time capture and visualization of
  verbal content in meetings}. In \bibinfo{booktitle}{\emph{Proceedings of the
  2019 CHI Conference on Human Factors in Computing Systems}}.
  \bibinfo{pages}{1--14}.
\newblock


\bibitem[\protect\citeauthoryear{Chang and Kang}{Chang and Kang}{2016}]%
        {chang2016challenges}
\bibfield{author}{\bibinfo{person}{Bo Chang} {and} \bibinfo{person}{Haijun
  Kang}.} \bibinfo{year}{2016}\natexlab{}.
\newblock \showarticletitle{Challenges facing group work online}.
\newblock \bibinfo{journal}{\emph{Distance Education}} \bibinfo{volume}{37},
  \bibinfo{number}{1} (\bibinfo{year}{2016}), \bibinfo{pages}{73--88}.
\newblock


\bibitem[\protect\citeauthoryear{Cheng, Smith-Renner, Zhang, Tetreault, and
  Jaimes}{Cheng et~al\mbox{.}}{2022}]%
        {cheng2022mapping}
\bibfield{author}{\bibinfo{person}{Ruijia Cheng}, \bibinfo{person}{Alison
  Smith-Renner}, \bibinfo{person}{Ke Zhang}, \bibinfo{person}{Joel~R
  Tetreault}, {and} \bibinfo{person}{Alejandro Jaimes}.}
  \bibinfo{year}{2022}\natexlab{}.
\newblock \showarticletitle{Mapping the Design Space of Human-AI Interaction in
  Text Summarization}.
\newblock \bibinfo{journal}{\emph{arXiv preprint arXiv:2206.14863}}
  (\bibinfo{year}{2022}).
\newblock


\bibitem[\protect\citeauthoryear{Chi and Wylie}{Chi and Wylie}{2014}]%
        {chi2014icap}
\bibfield{author}{\bibinfo{person}{Michelene~TH Chi} {and}
  \bibinfo{person}{Ruth Wylie}.} \bibinfo{year}{2014}\natexlab{}.
\newblock \showarticletitle{The ICAP framework: Linking cognitive engagement to
  active learning outcomes}.
\newblock \bibinfo{journal}{\emph{Educational psychologist}}
  \bibinfo{volume}{49}, \bibinfo{number}{4} (\bibinfo{year}{2014}),
  \bibinfo{pages}{219--243}.
\newblock


\bibitem[\protect\citeauthoryear{CHIA, GHAVIFEKR, and RAZAK}{CHIA
  et~al\mbox{.}}{[n.d.]}]%
        {chiaonline}
\bibfield{author}{\bibinfo{person}{CHI-KUAN CHIA}, \bibinfo{person}{SIMIN
  GHAVIFEKR}, {and} \bibinfo{person}{AHMAD ZABIDI BIN~ABDUL RAZAK}.}
  \bibinfo{year}{[n.d.]}\natexlab{}.
\newblock \showarticletitle{Online Interview Tools for Qualitative Data
  Collection During COVID-19 Pandemic: Review of Web Conferencing Platforms'
  Functionality}.
\newblock  (\bibinfo{year}{[n.\,d.]}).
\newblock


\bibitem[\protect\citeauthoryear{Cho, Im, Lee, and Lee}{Cho
  et~al\mbox{.}}{2021}]%
        {cho2021want}
\bibfield{author}{\bibinfo{person}{Haena Cho}, \bibinfo{person}{Hyeonjeong Im},
  \bibinfo{person}{Sunok Lee}, {and} \bibinfo{person}{Sangsu Lee}.}
  \bibinfo{year}{2021}\natexlab{}.
\newblock \showarticletitle{“I want more than” User-generated Icons for
  Better Video-mediated Communications on the Collaborative Design Process}. In
  \bibinfo{booktitle}{\emph{Extended Abstracts of the 2021 CHI Conference on
  Human Factors in Computing Systems}}. \bibinfo{pages}{1--6}.
\newblock


\bibitem[\protect\citeauthoryear{Cook}{Cook}{2009}]%
        {cook2009building}
\bibfield{author}{\bibinfo{person}{Sarah Cook}.}
  \bibinfo{year}{2009}\natexlab{}.
\newblock \bibinfo{booktitle}{\emph{Building a High Performance Team: Proven
  techniques for effective team working}}.
\newblock \bibinfo{publisher}{IT Governance Ltd}.
\newblock


\bibitem[\protect\citeauthoryear{Dennis and Kinney}{Dennis and Kinney}{1998}]%
        {dennis1998testing}
\bibfield{author}{\bibinfo{person}{Alan~R Dennis} {and}
  \bibinfo{person}{Susan~T Kinney}.} \bibinfo{year}{1998}\natexlab{}.
\newblock \showarticletitle{Testing media richness theory in the new media: The
  effects of cues, feedback, and task equivocality}.
\newblock \bibinfo{journal}{\emph{Information systems research}}
  \bibinfo{volume}{9}, \bibinfo{number}{3} (\bibinfo{year}{1998}),
  \bibinfo{pages}{256--274}.
\newblock


\bibitem[\protect\citeauthoryear{Dhawan, Carnes, Byars-Winston, and
  Duma}{Dhawan et~al\mbox{.}}{2021}]%
        {dhawan2021videoconferencing}
\bibfield{author}{\bibinfo{person}{Natasha Dhawan}, \bibinfo{person}{Molly
  Carnes}, \bibinfo{person}{Angela Byars-Winston}, {and}
  \bibinfo{person}{Narjust Duma}.} \bibinfo{year}{2021}\natexlab{}.
\newblock \showarticletitle{Videoconferencing etiquette: promoting gender
  equity during virtual meetings}.
\newblock \bibinfo{journal}{\emph{Journal of Women's Health}}
  \bibinfo{volume}{30}, \bibinfo{number}{4} (\bibinfo{year}{2021}),
  \bibinfo{pages}{460--465}.
\newblock


\bibitem[\protect\citeauthoryear{Dong and Fu}{Dong and Fu}{2012}]%
        {dong2012one}
\bibfield{author}{\bibinfo{person}{Wei Dong} {and} \bibinfo{person}{Wai-Tat
  Fu}.} \bibinfo{year}{2012}\natexlab{}.
\newblock \showarticletitle{One piece at a time: why video-based communication
  is better for negotiation and conflict resolution}. In
  \bibinfo{booktitle}{\emph{Proceedings of the ACM 2012 conference on Computer
  Supported Cooperative Work}}. \bibinfo{pages}{167--176}.
\newblock


\bibitem[\protect\citeauthoryear{Duan, Yamashita, Hwang, and Fussell}{Duan
  et~al\mbox{.}}{2018}]%
        {duan2018let}
\bibfield{author}{\bibinfo{person}{Wen Duan}, \bibinfo{person}{Naomi
  Yamashita}, \bibinfo{person}{Sun~Young Hwang}, {and} \bibinfo{person}{Susan
  Fussell}.} \bibinfo{year}{2018}\natexlab{}.
\newblock \showarticletitle{" Let Me Ask Them to Clarify If You Don't Want
  To"-A Clarification Agent for Nonnative Speakers}. In
  \bibinfo{booktitle}{\emph{Extended Abstracts of the 2018 CHI Conference on
  Human Factors in Computing Systems}}. \bibinfo{pages}{1--6}.
\newblock


\bibitem[\protect\citeauthoryear{Duan, Yamashita, Shirai, and Fussell}{Duan
  et~al\mbox{.}}{2021}]%
        {duan2021bridging}
\bibfield{author}{\bibinfo{person}{Wen Duan}, \bibinfo{person}{Naomi
  Yamashita}, \bibinfo{person}{Yoshinari Shirai}, {and}
  \bibinfo{person}{Susan~R Fussell}.} \bibinfo{year}{2021}\natexlab{}.
\newblock \showarticletitle{Bridging Fluency Disparity between Native and
  Nonnative Speakers in Multilingual Multiparty Collaboration Using a
  Clarification Agent}.
\newblock \bibinfo{journal}{\emph{Proceedings of the ACM on Human-Computer
  Interaction}} \bibinfo{volume}{5}, \bibinfo{number}{CSCW2}
  (\bibinfo{year}{2021}), \bibinfo{pages}{1--31}.
\newblock


\bibitem[\protect\citeauthoryear{El-Assady, Gold, Acevedo, Collins, and
  Keim}{El-Assady et~al\mbox{.}}{2016}]%
        {el2016contovi}
\bibfield{author}{\bibinfo{person}{Mennatallah El-Assady},
  \bibinfo{person}{Valentin Gold}, \bibinfo{person}{Carmela Acevedo},
  \bibinfo{person}{Christopher Collins}, {and} \bibinfo{person}{Daniel Keim}.}
  \bibinfo{year}{2016}\natexlab{}.
\newblock \showarticletitle{ConToVi: Multi-party conversation exploration using
  topic-space views}. In \bibinfo{booktitle}{\emph{Computer Graphics Forum}},
  Vol.~\bibinfo{volume}{35}. Wiley Online Library, \bibinfo{pages}{431--440}.
\newblock


\bibitem[\protect\citeauthoryear{Fahy}{Fahy}{2006}]%
        {fahy2006online}
\bibfield{author}{\bibinfo{person}{Patrick~J Fahy}.}
  \bibinfo{year}{2006}\natexlab{}.
\newblock \showarticletitle{Online and face-to-face group interaction processes
  compared using Bales' interaction process analysis (IPA)}.
\newblock \bibinfo{journal}{\emph{European Journal of open, distance and
  e-learning}} \bibinfo{volume}{9}, \bibinfo{number}{1} (\bibinfo{year}{2006}).
\newblock


\bibitem[\protect\citeauthoryear{Fang, Wang, Yang, Liu, and Wang}{Fang
  et~al\mbox{.}}{2022}]%
        {fang2022understanding}
\bibfield{author}{\bibinfo{person}{Jingchao Fang}, \bibinfo{person}{Yanhao
  Wang}, \bibinfo{person}{Chi-Lan Yang}, \bibinfo{person}{Ching Liu}, {and}
  \bibinfo{person}{Hao-Chuan Wang}.} \bibinfo{year}{2022}\natexlab{}.
\newblock \showarticletitle{Understanding the Effects of Structured Note-taking
  Systems for Video-based Learners in Individual and Social Learning Contexts}.
\newblock \bibinfo{journal}{\emph{Proceedings of the ACM on Human-Computer
  Interaction}} \bibinfo{volume}{6}, \bibinfo{number}{GROUP}
  (\bibinfo{year}{2022}), \bibinfo{pages}{1--21}.
\newblock


\bibitem[\protect\citeauthoryear{Fang, Wang, Yang, and Wang}{Fang
  et~al\mbox{.}}{2021}]%
        {fang2021notecostruct}
\bibfield{author}{\bibinfo{person}{Jingchao Fang}, \bibinfo{person}{Yanhao
  Wang}, \bibinfo{person}{Chi-Lan Yang}, {and} \bibinfo{person}{Hao-Chuan
  Wang}.} \bibinfo{year}{2021}\natexlab{}.
\newblock \showarticletitle{NoteCoStruct: Powering Online Learners with
  Socially Scaffolded Note Taking and Sharing}. In
  \bibinfo{booktitle}{\emph{Extended Abstracts of the 2021 CHI Conference on
  Human Factors in Computing Systems}}. \bibinfo{pages}{1--5}.
\newblock


\bibitem[\protect\citeauthoryear{Fanguy, Baldwin, Shmeleva, Lee, and
  Costley}{Fanguy et~al\mbox{.}}{2021}]%
        {fanguy2021collaboration}
\bibfield{author}{\bibinfo{person}{Mik Fanguy}, \bibinfo{person}{Matthew
  Baldwin}, \bibinfo{person}{Evgeniia Shmeleva}, \bibinfo{person}{Kyungmee
  Lee}, {and} \bibinfo{person}{Jamie Costley}.}
  \bibinfo{year}{2021}\natexlab{}.
\newblock \showarticletitle{How collaboration influences the effect of
  note-taking on writing performance and recall of contents}.
\newblock \bibinfo{journal}{\emph{Interactive Learning Environments}}
  (\bibinfo{year}{2021}), \bibinfo{pages}{1--15}.
\newblock


\bibitem[\protect\citeauthoryear{Fowler}{Fowler}{2014}]%
        {fowler2014talking}
\bibfield{author}{\bibinfo{person}{Robin Fowler}.}
  \bibinfo{year}{2014}\natexlab{}.
\newblock \showarticletitle{Talking teams: Increased equity in participation in
  online compared to face-to-face team discussions}. In
  \bibinfo{booktitle}{\emph{2014 ASEE Annual Conference \& Exposition}}.
  \bibinfo{pages}{24--1154}.
\newblock


\bibitem[\protect\citeauthoryear{Fu, Zhao, Cheng, Zhu, and Marlow}{Fu
  et~al\mbox{.}}{2018}]%
        {fu2018t}
\bibfield{author}{\bibinfo{person}{Siwei Fu}, \bibinfo{person}{Jian Zhao},
  \bibinfo{person}{Hao~Fei Cheng}, \bibinfo{person}{Haiyi Zhu}, {and}
  \bibinfo{person}{Jennifer Marlow}.} \bibinfo{year}{2018}\natexlab{}.
\newblock \showarticletitle{T-cal: Understanding team conversational data with
  calendar-based visualization}. In \bibinfo{booktitle}{\emph{Proceedings of
  the 2018 CHI Conference on Human Factors in Computing Systems}}.
  \bibinfo{pages}{1--13}.
\newblock


\bibitem[\protect\citeauthoryear{Gao, Yamashita, Hautasaari, Echenique, and
  Fussell}{Gao et~al\mbox{.}}{2014}]%
        {gao2014effects}
\bibfield{author}{\bibinfo{person}{Ge Gao}, \bibinfo{person}{Naomi Yamashita},
  \bibinfo{person}{Ari~MJ Hautasaari}, \bibinfo{person}{Andy Echenique}, {and}
  \bibinfo{person}{Susan~R Fussell}.} \bibinfo{year}{2014}\natexlab{}.
\newblock \showarticletitle{Effects of public vs. private automated transcripts
  on multiparty communication between native and non-native English speakers}.
  In \bibinfo{booktitle}{\emph{Proceedings of the SIGCHI Conference on Human
  Factors in Computing Systems}}. \bibinfo{pages}{843--852}.
\newblock


\bibitem[\protect\citeauthoryear{Geller, Hoernle, Gal, Segal, Zhang, Karger,
  Facciotti, and Igo}{Geller et~al\mbox{.}}{2020}]%
        {geller2020confused}
\bibfield{author}{\bibinfo{person}{Shay~A Geller}, \bibinfo{person}{Nicholas
  Hoernle}, \bibinfo{person}{Kobi Gal}, \bibinfo{person}{Avi Segal},
  \bibinfo{person}{Amy~X Zhang}, \bibinfo{person}{David Karger},
  \bibinfo{person}{Marc~T Facciotti}, {and} \bibinfo{person}{Michele Igo}.}
  \bibinfo{year}{2020}\natexlab{}.
\newblock \showarticletitle{\# Confused and beyond: detecting confusion in
  course forums using students' hashtags}. In
  \bibinfo{booktitle}{\emph{Proceedings of the Tenth International Conference
  on Learning Analytics \& Knowledge}}. \bibinfo{pages}{589--594}.
\newblock


\bibitem[\protect\citeauthoryear{Green and Lazarus}{Green and Lazarus}{1991}]%
        {green1991today}
\bibfield{author}{\bibinfo{person}{Walter~A Green} {and}
  \bibinfo{person}{Harold Lazarus}.} \bibinfo{year}{1991}\natexlab{}.
\newblock \showarticletitle{Are Today's Executives Meeting with Success?}
\newblock \bibinfo{journal}{\emph{Journal of management development}}
  (\bibinfo{year}{1991}).
\newblock


\bibitem[\protect\citeauthoryear{Grgurovi{\'c} and Hegelheimer}{Grgurovi{\'c}
  and Hegelheimer}{2007}]%
        {grgurovic2007help}
\bibfield{author}{\bibinfo{person}{Maja Grgurovi{\'c}} {and}
  \bibinfo{person}{Volker Hegelheimer}.} \bibinfo{year}{2007}\natexlab{}.
\newblock \showarticletitle{Help options and multimedia listening: Students'
  use of subtitles and the transcript}.
\newblock \bibinfo{journal}{\emph{Language learning \& technology}}
  \bibinfo{volume}{11}, \bibinfo{number}{1} (\bibinfo{year}{2007}),
  \bibinfo{pages}{45--66}.
\newblock


\bibitem[\protect\citeauthoryear{He, Xiong, and Xia}{He et~al\mbox{.}}{2021}]%
        {he2021you}
\bibfield{author}{\bibinfo{person}{Muchen He}, \bibinfo{person}{Beibei Xiong},
  {and} \bibinfo{person}{Kaseya Xia}.} \bibinfo{year}{2021}\natexlab{}.
\newblock \showarticletitle{Are You Looking at Me? Eye Gazing in Web Video
  Conferences}.
\newblock \bibinfo{journal}{\emph{methods}}  \bibinfo{volume}{27}
  (\bibinfo{year}{2021}), \bibinfo{pages}{28}.
\newblock


\bibitem[\protect\citeauthoryear{Heath and Bryant}{Heath and Bryant}{2013}]%
        {heath2013human}
\bibfield{author}{\bibinfo{person}{Robert~L Heath} {and}
  \bibinfo{person}{Jennings Bryant}.} \bibinfo{year}{2013}\natexlab{}.
\newblock \bibinfo{booktitle}{\emph{Human communication theory and research:
  Concepts, contexts, and challenges}}.
\newblock \bibinfo{publisher}{Routledge}.
\newblock


\bibitem[\protect\citeauthoryear{Hinze, Bischoff, and Blakowski}{Hinze
  et~al\mbox{.}}{2002}]%
        {hinze2002jigsaw}
\bibfield{author}{\bibinfo{person}{Udo Hinze}, \bibinfo{person}{Michael
  Bischoff}, {and} \bibinfo{person}{Gerold Blakowski}.}
  \bibinfo{year}{2002}\natexlab{}.
\newblock \showarticletitle{Jigsaw Method in the Context of CSCL}. In
  \bibinfo{booktitle}{\emph{EdMedia+ Innovate Learning}}. Association for the
  Advancement of Computing in Education (AACE), \bibinfo{pages}{789--794}.
\newblock


\bibitem[\protect\citeauthoryear{Hong, Conroy, Favre, Kulesza, Lin, and
  Nenkova}{Hong et~al\mbox{.}}{2014}]%
        {hong2014repository}
\bibfield{author}{\bibinfo{person}{Kai Hong}, \bibinfo{person}{John Conroy},
  \bibinfo{person}{Benoit Favre}, \bibinfo{person}{Alex Kulesza},
  \bibinfo{person}{Hui Lin}, {and} \bibinfo{person}{Ani Nenkova}.}
  \bibinfo{year}{2014}\natexlab{}.
\newblock \showarticletitle{A repository of state of the art and competitive
  baseline summaries for generic news summarization}. In
  \bibinfo{booktitle}{\emph{Proceedings of the Ninth International Conference
  on Language Resources and Evaluation (LREC'14)}}.
  \bibinfo{pages}{1608--1616}.
\newblock


\bibitem[\protect\citeauthoryear{Huber, Shin, Russell, Wang, and Mysore}{Huber
  et~al\mbox{.}}{2019}]%
        {huber2019b}
\bibfield{author}{\bibinfo{person}{Bernd Huber},
  \bibinfo{person}{Hijung~Valentina Shin}, \bibinfo{person}{Bryan Russell},
  \bibinfo{person}{Oliver Wang}, {and} \bibinfo{person}{Gautham~J Mysore}.}
  \bibinfo{year}{2019}\natexlab{}.
\newblock \showarticletitle{B-script: Transcript-based B-roll video editing
  with recommendations}. In \bibinfo{booktitle}{\emph{Proceedings of the 2019
  CHI Conference on Human Factors in Computing Systems}}.
  \bibinfo{pages}{1--11}.
\newblock


\bibitem[\protect\citeauthoryear{Jonassen and Kwon}{Jonassen and Kwon}{2001}]%
        {jonassen2001communication}
\bibfield{author}{\bibinfo{person}{David~H Jonassen} {and}
  \bibinfo{person}{Hyug Kwon}.} \bibinfo{year}{2001}\natexlab{}.
\newblock \showarticletitle{Communication patterns in computer mediated versus
  face-to-face group problem solving}.
\newblock \bibinfo{journal}{\emph{Educational technology research and
  development}} \bibinfo{volume}{49}, \bibinfo{number}{1}
  (\bibinfo{year}{2001}), \bibinfo{pages}{35}.
\newblock


\bibitem[\protect\citeauthoryear{Jones}{Jones}{1999}]%
        {jones1999silence}
\bibfield{author}{\bibinfo{person}{Jeremy~F Jones}.}
  \bibinfo{year}{1999}\natexlab{}.
\newblock \showarticletitle{From Silence to Talk: Cross-Cultural Ideas
  onStudents Participation in Academic Group Discussion}.
\newblock \bibinfo{journal}{\emph{English for specific Purposes}}
  \bibinfo{volume}{18}, \bibinfo{number}{3} (\bibinfo{year}{1999}),
  \bibinfo{pages}{243--259}.
\newblock


\bibitem[\protect\citeauthoryear{Junuzovic, Inkpen, Hegde, Zhang, Tang, and
  Brooks}{Junuzovic et~al\mbox{.}}{2011}]%
        {junuzovic2011did}
\bibfield{author}{\bibinfo{person}{Sasa Junuzovic}, \bibinfo{person}{Kori
  Inkpen}, \bibinfo{person}{Rajesh Hegde}, \bibinfo{person}{Zhengyou Zhang},
  \bibinfo{person}{John Tang}, {and} \bibinfo{person}{Christopher Brooks}.}
  \bibinfo{year}{2011}\natexlab{}.
\newblock \showarticletitle{What did I miss? In-meeting review using multimodal
  accelerated instant replay (AIR) conferencing}. In
  \bibinfo{booktitle}{\emph{Proceedings of the SIGCHI Conference on Human
  Factors in Computing Systems}}. \bibinfo{pages}{513--522}.
\newblock


\bibitem[\protect\citeauthoryear{Kafle, Dingman, and Huenerfauth}{Kafle
  et~al\mbox{.}}{2021}]%
        {kafle2021deaf}
\bibfield{author}{\bibinfo{person}{Sushant Kafle}, \bibinfo{person}{Becca
  Dingman}, {and} \bibinfo{person}{Matt Huenerfauth}.}
  \bibinfo{year}{2021}\natexlab{}.
\newblock \showarticletitle{Deaf and Hard-of-hearing Users Evaluating Designs
  for Highlighting Key Words in Educational Lecture Videos}.
\newblock \bibinfo{journal}{\emph{ACM Transactions on Accessible Computing
  (TACCESS)}} \bibinfo{volume}{14}, \bibinfo{number}{4} (\bibinfo{year}{2021}),
  \bibinfo{pages}{1--24}.
\newblock


\bibitem[\protect\citeauthoryear{Kam, Wang, Iles, Tse, Chiu, Glaser, Tarshish,
  and Canny}{Kam et~al\mbox{.}}{2005}]%
        {kam2005livenotes}
\bibfield{author}{\bibinfo{person}{Matthew Kam}, \bibinfo{person}{Jingtao
  Wang}, \bibinfo{person}{Alastair Iles}, \bibinfo{person}{Eric Tse},
  \bibinfo{person}{Jane Chiu}, \bibinfo{person}{Daniel Glaser},
  \bibinfo{person}{Orna Tarshish}, {and} \bibinfo{person}{John Canny}.}
  \bibinfo{year}{2005}\natexlab{}.
\newblock \showarticletitle{Livenotes: a system for cooperative and augmented
  note-taking in lectures}. In \bibinfo{booktitle}{\emph{Proceedings of the
  SIGCHI conference on Human factors in computing systems}}.
  \bibinfo{pages}{531--540}.
\newblock


\bibitem[\protect\citeauthoryear{Karl, Peluchette, and Aghakhani}{Karl
  et~al\mbox{.}}{2021}]%
        {karl2021virtual}
\bibfield{author}{\bibinfo{person}{Katherine~A Karl}, \bibinfo{person}{Joy~V
  Peluchette}, {and} \bibinfo{person}{Navid Aghakhani}.}
  \bibinfo{year}{2021}\natexlab{}.
\newblock \showarticletitle{Virtual Work Meetings During the COVID-19 Pandemic:
  The Good, Bad, and Ugly}.
\newblock \bibinfo{journal}{\emph{Small Group Research}}
  (\bibinfo{year}{2021}), \bibinfo{pages}{10464964211015286}.
\newblock


\bibitem[\protect\citeauthoryear{Kim, Guo, Cai, Li, Gajos, and Miller}{Kim
  et~al\mbox{.}}{2014}]%
        {kim2014data}
\bibfield{author}{\bibinfo{person}{Juho Kim}, \bibinfo{person}{Philip~J Guo},
  \bibinfo{person}{Carrie~J Cai}, \bibinfo{person}{Shang-Wen Li},
  \bibinfo{person}{Krzysztof~Z Gajos}, {and} \bibinfo{person}{Robert~C
  Miller}.} \bibinfo{year}{2014}\natexlab{}.
\newblock \showarticletitle{Data-driven interaction techniques for improving
  navigation of educational videos}. In \bibinfo{booktitle}{\emph{Proceedings
  of the 27th annual ACM symposium on User interface software and technology}}.
  \bibinfo{pages}{563--572}.
\newblock


\bibitem[\protect\citeauthoryear{Kirshenbaum, Davidson, Harden, North,
  Kobayashi, Theriot, Tabalba~Jr, Rogers, Belcaid, Burks,
  et~al\mbox{.}}{Kirshenbaum et~al\mbox{.}}{2021}]%
        {kirshenbaum2021traces}
\bibfield{author}{\bibinfo{person}{Nurit Kirshenbaum}, \bibinfo{person}{Kylie
  Davidson}, \bibinfo{person}{Jesse Harden}, \bibinfo{person}{Chris North},
  \bibinfo{person}{Dylan Kobayashi}, \bibinfo{person}{Ryan Theriot},
  \bibinfo{person}{Roderick~S Tabalba~Jr}, \bibinfo{person}{Michael~L Rogers},
  \bibinfo{person}{Mahdi Belcaid}, \bibinfo{person}{Andrew~T Burks},
  {et~al\mbox{.}}} \bibinfo{year}{2021}\natexlab{}.
\newblock \showarticletitle{Traces of Time through Space: Advantages of
  Creating Complex Canvases in Collaborative Meetings}.
\newblock \bibinfo{journal}{\emph{Proceedings of the ACM on Human-Computer
  Interaction}} \bibinfo{volume}{5}, \bibinfo{number}{ISS}
  (\bibinfo{year}{2021}), \bibinfo{pages}{1--20}.
\newblock


\bibitem[\protect\citeauthoryear{Kohnke and Moorhouse}{Kohnke and
  Moorhouse}{2020}]%
        {kohnke2020facilitating}
\bibfield{author}{\bibinfo{person}{Lucas Kohnke} {and}
  \bibinfo{person}{Benjamin~Luke Moorhouse}.} \bibinfo{year}{2020}\natexlab{}.
\newblock \showarticletitle{Facilitating synchronous online language learning
  through Zoom}.
\newblock \bibinfo{journal}{\emph{RELC Journal}} (\bibinfo{year}{2020}),
  \bibinfo{pages}{0033688220937235}.
\newblock


\bibitem[\protect\citeauthoryear{Kuzminykh and Rintel}{Kuzminykh and
  Rintel}{2020}]%
        {kuzminykh2020low}
\bibfield{author}{\bibinfo{person}{Anastasia Kuzminykh} {and}
  \bibinfo{person}{Sean Rintel}.} \bibinfo{year}{2020}\natexlab{}.
\newblock \showarticletitle{Low Engagement As a Deliberate Practice of Remote
  Participants in Video Meetings}. In \bibinfo{booktitle}{\emph{Extended
  Abstracts of the 2020 CHI Conference on Human Factors in Computing Systems}}.
  \bibinfo{pages}{1--9}.
\newblock


\bibitem[\protect\citeauthoryear{Lampe, Johnston, and Resnick}{Lampe
  et~al\mbox{.}}{2007}]%
        {lampe2007follow}
\bibfield{author}{\bibinfo{person}{Cliff~AC Lampe}, \bibinfo{person}{Erik
  Johnston}, {and} \bibinfo{person}{Paul Resnick}.}
  \bibinfo{year}{2007}\natexlab{}.
\newblock \showarticletitle{Follow the reader: filtering comments on slashdot}.
  In \bibinfo{booktitle}{\emph{Proceedings of the SIGCHI conference on Human
  factors in computing systems}}. \bibinfo{pages}{1253--1262}.
\newblock


\bibitem[\protect\citeauthoryear{Larsson and Wahlgren}{Larsson and
  Wahlgren}{2020}]%
        {larsson2020communication}
\bibfield{author}{\bibinfo{person}{Daniella Larsson} {and}
  \bibinfo{person}{Matilda Wahlgren}.} \bibinfo{year}{2020}\natexlab{}.
\newblock \bibinfo{title}{Communication challenges perceived by leaders of
  different sized virtual teams, and how they are managed: Experiences from
  leaders within Swedish organizations}.
\newblock
\newblock


\bibitem[\protect\citeauthoryear{Lins, Oliveira, Cabral, Batista, Tenorio,
  Ferreira, Lima, de~Fran{\c{c}}a Pereira~e Silva, and Simske}{Lins
  et~al\mbox{.}}{2019}]%
        {lins2019cnn}
\bibfield{author}{\bibinfo{person}{Rafael~Dueire Lins},
  \bibinfo{person}{Hilario Oliveira}, \bibinfo{person}{Luciano Cabral},
  \bibinfo{person}{Jamilson Batista}, \bibinfo{person}{Bruno Tenorio},
  \bibinfo{person}{Rafael Ferreira}, \bibinfo{person}{Rinaldo Lima},
  \bibinfo{person}{Gabriel de~Fran{\c{c}}a Pereira~e Silva}, {and}
  \bibinfo{person}{Steven~J Simske}.} \bibinfo{year}{2019}\natexlab{}.
\newblock \showarticletitle{The cnn-corpus: A large textual corpus for
  single-document extractive summarization}. In
  \bibinfo{booktitle}{\emph{Proceedings of the ACM Symposium on Document
  Engineering 2019}}. \bibinfo{pages}{1--10}.
\newblock


\bibitem[\protect\citeauthoryear{Lucero}{Lucero}{2015}]%
        {lucero2015using}
\bibfield{author}{\bibinfo{person}{Andr{\'e}s Lucero}.}
  \bibinfo{year}{2015}\natexlab{}.
\newblock \showarticletitle{Using affinity diagrams to evaluate interactive
  prototypes}. In \bibinfo{booktitle}{\emph{IFIP Conference on Human-Computer
  Interaction}}. Springer, \bibinfo{pages}{231--248}.
\newblock


\bibitem[\protect\citeauthoryear{Lunenburg}{Lunenburg}{2010}]%
        {lunenburg2010communication}
\bibfield{author}{\bibinfo{person}{Fred~C Lunenburg}.}
  \bibinfo{year}{2010}\natexlab{}.
\newblock \showarticletitle{Communication: The process, barriers, and improving
  effectiveness}.
\newblock \bibinfo{journal}{\emph{Schooling}} \bibinfo{volume}{1},
  \bibinfo{number}{1} (\bibinfo{year}{2010}), \bibinfo{pages}{1--10}.
\newblock


\bibitem[\protect\citeauthoryear{Luria, Zheng, Huffman, Huang, Zimmerman, and
  Forlizzi}{Luria et~al\mbox{.}}{2020}]%
        {luria2020social}
\bibfield{author}{\bibinfo{person}{Michal Luria}, \bibinfo{person}{Rebecca
  Zheng}, \bibinfo{person}{Bennett Huffman}, \bibinfo{person}{Shuangni Huang},
  \bibinfo{person}{John Zimmerman}, {and} \bibinfo{person}{Jodi Forlizzi}.}
  \bibinfo{year}{2020}\natexlab{}.
\newblock \showarticletitle{Social Boundaries for Personal Agents in the
  Interpersonal Space of the Home}. In \bibinfo{booktitle}{\emph{Proceedings of
  the 2020 CHI Conference on Human Factors in Computing Systems}}.
  \bibinfo{pages}{1--12}.
\newblock


\bibitem[\protect\citeauthoryear{Masek, Ismail, Hashim, and Mohd}{Masek
  et~al\mbox{.}}{2021}]%
        {masek2021defining}
\bibfield{author}{\bibinfo{person}{Alias Masek}, \bibinfo{person}{Affero
  Ismail}, \bibinfo{person}{Suhaizal Hashim}, {and}
  \bibinfo{person}{Salsabella~Fizol Mohd}.} \bibinfo{year}{2021}\natexlab{}.
\newblock \showarticletitle{Defining students’ active participation in a
  group discussion session from different perspectives}.
\newblock \bibinfo{journal}{\emph{Academia}} \bibinfo{number}{23-24}
  (\bibinfo{year}{2021}), \bibinfo{pages}{67--84}.
\newblock


\bibitem[\protect\citeauthoryear{Miniukovich, Scaltritti, Sulpizio, and
  De~Angeli}{Miniukovich et~al\mbox{.}}{2019}]%
        {miniukovich2019guideline}
\bibfield{author}{\bibinfo{person}{Aliaksei Miniukovich},
  \bibinfo{person}{Michele Scaltritti}, \bibinfo{person}{Simone Sulpizio},
  {and} \bibinfo{person}{Antonella De~Angeli}.}
  \bibinfo{year}{2019}\natexlab{}.
\newblock \showarticletitle{Guideline-based evaluation of web readability}. In
  \bibinfo{booktitle}{\emph{Proceedings of the 2019 CHI Conference on Human
  Factors in Computing Systems}}. \bibinfo{pages}{1--12}.
\newblock


\bibitem[\protect\citeauthoryear{Murali, Hernandez, McDuff, Rowan, Suh, and
  Czerwinski}{Murali et~al\mbox{.}}{2021}]%
        {murali2021affectivespotlight}
\bibfield{author}{\bibinfo{person}{Prasanth Murali}, \bibinfo{person}{Javier
  Hernandez}, \bibinfo{person}{Daniel McDuff}, \bibinfo{person}{Kael Rowan},
  \bibinfo{person}{Jina Suh}, {and} \bibinfo{person}{Mary Czerwinski}.}
  \bibinfo{year}{2021}\natexlab{}.
\newblock \showarticletitle{AffectiveSpotlight: Facilitating the Communication
  of Affective Responses from Audience Members during Online Presentations}. In
  \bibinfo{booktitle}{\emph{Proceedings of the 2021 CHI Conference on Human
  Factors in Computing Systems}}. \bibinfo{pages}{1--13}.
\newblock


\bibitem[\protect\citeauthoryear{Myers and Anderson}{Myers and
  Anderson}{2008}]%
        {myers2008fundamentals}
\bibfield{author}{\bibinfo{person}{Scott~A Myers} {and}
  \bibinfo{person}{Carolyn~M Anderson}.} \bibinfo{year}{2008}\natexlab{}.
\newblock \bibinfo{booktitle}{\emph{The fundamentals of small group
  communication}}.
\newblock \bibinfo{publisher}{Sage Publications}.
\newblock


\bibitem[\protect\citeauthoryear{Namikawa, Suzuki, Iijima, Sarcar, and
  Ochiai}{Namikawa et~al\mbox{.}}{2021}]%
        {namikawa2021emojicam}
\bibfield{author}{\bibinfo{person}{Kosaku Namikawa}, \bibinfo{person}{Ippei
  Suzuki}, \bibinfo{person}{Ryo Iijima}, \bibinfo{person}{Sayan Sarcar}, {and}
  \bibinfo{person}{Yoichi Ochiai}.} \bibinfo{year}{2021}\natexlab{}.
\newblock \showarticletitle{EmojiCam: Emoji-Assisted Video Communication System
  Leveraging Facial Expressions}. In \bibinfo{booktitle}{\emph{International
  Conference on Human-Computer Interaction}}. Springer,
  \bibinfo{pages}{611--625}.
\newblock


\bibitem[\protect\citeauthoryear{Nelson and Schunn}{Nelson and Schunn}{2009}]%
        {nelson2009nature}
\bibfield{author}{\bibinfo{person}{Melissa~M Nelson} {and}
  \bibinfo{person}{Christian~D Schunn}.} \bibinfo{year}{2009}\natexlab{}.
\newblock \showarticletitle{The nature of feedback: How different types of peer
  feedback affect writing performance}.
\newblock \bibinfo{journal}{\emph{Instructional science}}  \bibinfo{volume}{37}
  (\bibinfo{year}{2009}), \bibinfo{pages}{375--401}.
\newblock


\bibitem[\protect\citeauthoryear{Pan, Yamashita, and Wang}{Pan
  et~al\mbox{.}}{2017}]%
        {pan2017task}
\bibfield{author}{\bibinfo{person}{Mei-Hua Pan}, \bibinfo{person}{Naomi
  Yamashita}, {and} \bibinfo{person}{Hao-Chuan Wang}.}
  \bibinfo{year}{2017}\natexlab{}.
\newblock \showarticletitle{Task rebalancing: Improving multilingual
  communication with native speakers-generated highlights on automated
  transcripts}. In \bibinfo{booktitle}{\emph{Proceedings of the 2017 ACM
  Conference on Computer Supported Cooperative Work and Social Computing}}.
  \bibinfo{pages}{310--321}.
\newblock


\bibitem[\protect\citeauthoryear{Pavel, Goldman, Hartmann, and Agrawala}{Pavel
  et~al\mbox{.}}{2015}]%
        {pavel2015sceneskim}
\bibfield{author}{\bibinfo{person}{Amy Pavel}, \bibinfo{person}{Dan~B Goldman},
  \bibinfo{person}{Bj{\"o}rn Hartmann}, {and} \bibinfo{person}{Maneesh
  Agrawala}.} \bibinfo{year}{2015}\natexlab{}.
\newblock \showarticletitle{Sceneskim: Searching and browsing movies using
  synchronized captions, scripts and plot summaries}. In
  \bibinfo{booktitle}{\emph{Proceedings of the 28th Annual ACM Symposium on
  User Interface Software \& Technology}}. \bibinfo{pages}{181--190}.
\newblock


\bibitem[\protect\citeauthoryear{Pavel, Reed, Hartmann, and Agrawala}{Pavel
  et~al\mbox{.}}{2014}]%
        {pavel2014video}
\bibfield{author}{\bibinfo{person}{Amy Pavel}, \bibinfo{person}{Colorado Reed},
  \bibinfo{person}{Bj{\"o}rn Hartmann}, {and} \bibinfo{person}{Maneesh
  Agrawala}.} \bibinfo{year}{2014}\natexlab{}.
\newblock \showarticletitle{Video digests: a browsable, skimmable format for
  informational lecture videos}. In \bibinfo{booktitle}{\emph{Proceedings of
  the 27th annual ACM symposium on User interface software and technology}}.
  \bibinfo{pages}{573--582}.
\newblock


\bibitem[\protect\citeauthoryear{Piolat, Olive, and Kellogg}{Piolat
  et~al\mbox{.}}{2005}]%
        {piolat2005cognitive}
\bibfield{author}{\bibinfo{person}{Annie Piolat}, \bibinfo{person}{Thierry
  Olive}, {and} \bibinfo{person}{Ronald~T Kellogg}.}
  \bibinfo{year}{2005}\natexlab{}.
\newblock \showarticletitle{Cognitive effort during note taking}.
\newblock \bibinfo{journal}{\emph{Applied cognitive psychology}}
  \bibinfo{volume}{19}, \bibinfo{number}{3} (\bibinfo{year}{2005}),
  \bibinfo{pages}{291--312}.
\newblock


\bibitem[\protect\citeauthoryear{Robertson}{Robertson}{2005}]%
        {robertson2005active}
\bibfield{author}{\bibinfo{person}{Kathryn Robertson}.}
  \bibinfo{year}{2005}\natexlab{}.
\newblock \showarticletitle{Active listening: more than just paying attention}.
\newblock \bibinfo{journal}{\emph{Australian family physician}}
  \bibinfo{volume}{34}, \bibinfo{number}{12} (\bibinfo{year}{2005}).
\newblock


\bibitem[\protect\citeauthoryear{Ruan, Willis, Xu, Davis, Jiang, Brunskill, and
  Landay}{Ruan et~al\mbox{.}}{2019}]%
        {ruan2019bookbuddy}
\bibfield{author}{\bibinfo{person}{Sherry Ruan}, \bibinfo{person}{Angelica
  Willis}, \bibinfo{person}{Qianyao Xu}, \bibinfo{person}{Glenn~M Davis},
  \bibinfo{person}{Liwei Jiang}, \bibinfo{person}{Emma Brunskill}, {and}
  \bibinfo{person}{James~A Landay}.} \bibinfo{year}{2019}\natexlab{}.
\newblock \showarticletitle{Bookbuddy: Turning digital materials into
  interactive foreign language lessons through a voice chatbot}. In
  \bibinfo{booktitle}{\emph{Proceedings of the Sixth (2019) ACM Conference on
  Learning@ Scale}}. \bibinfo{pages}{1--4}.
\newblock


\bibitem[\protect\citeauthoryear{Samrose, McDuff, Sim, Suh, Rowan, Hernandez,
  Rintel, Moynihan, and Czerwinski}{Samrose et~al\mbox{.}}{2021}]%
        {samrose2021meetingcoach}
\bibfield{author}{\bibinfo{person}{Samiha Samrose}, \bibinfo{person}{Daniel
  McDuff}, \bibinfo{person}{Robert Sim}, \bibinfo{person}{Jina Suh},
  \bibinfo{person}{Kael Rowan}, \bibinfo{person}{Javier Hernandez},
  \bibinfo{person}{Sean Rintel}, \bibinfo{person}{Kevin Moynihan}, {and}
  \bibinfo{person}{Mary Czerwinski}.} \bibinfo{year}{2021}\natexlab{}.
\newblock \showarticletitle{MeetingCoach: An Intelligent Dashboard for
  Supporting Effective \& Inclusive Meetings}. In \bibinfo{booktitle}{\emph{CHI
  2021}}.
\newblock
\urldef\tempurl%
\url{https://www.microsoft.com/en-us/research/publication/meetingcoach-an-intelligent-dashboard-for-supporting-effective-inclusive-meetings/}
\showURL{%
\tempurl}


\bibitem[\protect\citeauthoryear{Samrose, Zhao, White, Li, Nova, Lu, Ali, and
  Hoque}{Samrose et~al\mbox{.}}{2018}]%
        {samrose2018coco}
\bibfield{author}{\bibinfo{person}{Samiha Samrose}, \bibinfo{person}{Ru Zhao},
  \bibinfo{person}{Jeffery White}, \bibinfo{person}{Vivian Li},
  \bibinfo{person}{Luis Nova}, \bibinfo{person}{Yichen Lu},
  \bibinfo{person}{Mohammad~Rafayet Ali}, {and} \bibinfo{person}{Mohammed~Ehsan
  Hoque}.} \bibinfo{year}{2018}\natexlab{}.
\newblock \showarticletitle{Coco: Collaboration coach for understanding team
  dynamics during video conferencing}.
\newblock \bibinfo{journal}{\emph{Proceedings of the ACM on interactive,
  mobile, wearable and ubiquitous technologies}} \bibinfo{volume}{1},
  \bibinfo{number}{4} (\bibinfo{year}{2018}), \bibinfo{pages}{1--24}.
\newblock


\bibitem[\protect\citeauthoryear{Sarkar, Rintel, Borowiec, Bergmann, Gillett,
  Bragg, Baym, and Sellen}{Sarkar et~al\mbox{.}}{2021}]%
        {sarkar2021promise}
\bibfield{author}{\bibinfo{person}{Advait Sarkar}, \bibinfo{person}{Sean
  Rintel}, \bibinfo{person}{Damian Borowiec}, \bibinfo{person}{Rachel
  Bergmann}, \bibinfo{person}{Sharon Gillett}, \bibinfo{person}{Danielle
  Bragg}, \bibinfo{person}{Nancy Baym}, {and} \bibinfo{person}{Abigail
  Sellen}.} \bibinfo{year}{2021}\natexlab{}.
\newblock \showarticletitle{The promise and peril of parallel chat in video
  meetings for work}. In \bibinfo{booktitle}{\emph{Extended Abstracts of the
  2021 CHI Conference on Human Factors in Computing Systems}}.
  \bibinfo{pages}{1--8}.
\newblock


\bibitem[\protect\citeauthoryear{Seita, Andrew, and Huenerfauth}{Seita
  et~al\mbox{.}}{2021}]%
        {seita2021deaf}
\bibfield{author}{\bibinfo{person}{Matthew Seita}, \bibinfo{person}{Sarah
  Andrew}, {and} \bibinfo{person}{Matt Huenerfauth}.}
  \bibinfo{year}{2021}\natexlab{}.
\newblock \showarticletitle{Deaf and hard-of-hearing users' preferences for
  hearing speakers' behavior during technology-mediated in-person and remote
  conversations}. In \bibinfo{booktitle}{\emph{Proceedings of the 18th
  International Web for All Conference}}. \bibinfo{pages}{1--12}.
\newblock


\bibitem[\protect\citeauthoryear{Seuren, Wherton, Greenhalgh, and Shaw}{Seuren
  et~al\mbox{.}}{2021}]%
        {seuren2021whose}
\bibfield{author}{\bibinfo{person}{Lucas~M Seuren}, \bibinfo{person}{Joseph
  Wherton}, \bibinfo{person}{Trisha Greenhalgh}, {and} \bibinfo{person}{Sara~E
  Shaw}.} \bibinfo{year}{2021}\natexlab{}.
\newblock \showarticletitle{Whose turn is it anyway? Latency and the
  organization of turn-taking in video-mediated interaction}.
\newblock \bibinfo{journal}{\emph{Journal of pragmatics}}
  \bibinfo{volume}{172} (\bibinfo{year}{2021}), \bibinfo{pages}{63--78}.
\newblock


\bibitem[\protect\citeauthoryear{Shannon, Hammer, Thurston, Diehl, and
  Dow}{Shannon et~al\mbox{.}}{2016}]%
        {shannon2016peerpresents}
\bibfield{author}{\bibinfo{person}{Amy Shannon}, \bibinfo{person}{Jessica
  Hammer}, \bibinfo{person}{Hassler Thurston}, \bibinfo{person}{Natalie Diehl},
  {and} \bibinfo{person}{Steven Dow}.} \bibinfo{year}{2016}\natexlab{}.
\newblock \showarticletitle{PeerPresents: A web-based system for in-class peer
  feedback during student presentations}. In
  \bibinfo{booktitle}{\emph{Proceedings of the 2016 ACM Conference on Designing
  Interactive Systems}}. \bibinfo{pages}{447--458}.
\newblock


\bibitem[\protect\citeauthoryear{Shi, Bryan, Bhamidipati, Zhao, Zhang, and
  Ma}{Shi et~al\mbox{.}}{2018}]%
        {shi2018meetingvis}
\bibfield{author}{\bibinfo{person}{Yang Shi}, \bibinfo{person}{Chris Bryan},
  \bibinfo{person}{Sridatt Bhamidipati}, \bibinfo{person}{Ying Zhao},
  \bibinfo{person}{Yaoxue Zhang}, {and} \bibinfo{person}{Kwan-Liu Ma}.}
  \bibinfo{year}{2018}\natexlab{}.
\newblock \showarticletitle{Meetingvis: Visual narratives to assist in
  recalling meeting context and content}.
\newblock \bibinfo{journal}{\emph{IEEE Transactions on Visualization and
  Computer Graphics}} \bibinfo{volume}{24}, \bibinfo{number}{6}
  (\bibinfo{year}{2018}), \bibinfo{pages}{1918--1929}.
\newblock


\bibitem[\protect\citeauthoryear{Stasser and Titus}{Stasser and Titus}{1985}]%
        {stasser1985pooling}
\bibfield{author}{\bibinfo{person}{Garold Stasser} {and}
  \bibinfo{person}{William Titus}.} \bibinfo{year}{1985}\natexlab{}.
\newblock \showarticletitle{Pooling of unshared information in group decision
  making: Biased information sampling during discussion.}
\newblock \bibinfo{journal}{\emph{Journal of personality and social
  psychology}} \bibinfo{volume}{48}, \bibinfo{number}{6}
  (\bibinfo{year}{1985}), \bibinfo{pages}{1467}.
\newblock


\bibitem[\protect\citeauthoryear{Stewart, Amon, Duran, and D'Mello}{Stewart
  et~al\mbox{.}}{2020}]%
        {stewart2020beyond}
\bibfield{author}{\bibinfo{person}{Angela~EB Stewart},
  \bibinfo{person}{Mary~Jean Amon}, \bibinfo{person}{Nicholas~D Duran}, {and}
  \bibinfo{person}{Sidney~K D'Mello}.} \bibinfo{year}{2020}\natexlab{}.
\newblock \showarticletitle{Beyond team makeup: Diversity in teams predicts
  valued outcomes in Computer-Mediated collaborations}. In
  \bibinfo{booktitle}{\emph{Proceedings of the 2020 CHI Conference on Human
  Factors in Computing Systems}}. \bibinfo{pages}{1--13}.
\newblock


\bibitem[\protect\citeauthoryear{Straus}{Straus}{1996}]%
        {straus1996getting}
\bibfield{author}{\bibinfo{person}{Susan~G Straus}.}
  \bibinfo{year}{1996}\natexlab{}.
\newblock \showarticletitle{Getting a clue: The effects of communication media
  and information distribution on participation and performance in
  computer-mediated and face-to-face groups}.
\newblock \bibinfo{journal}{\emph{Small group research}} \bibinfo{volume}{27},
  \bibinfo{number}{1} (\bibinfo{year}{1996}), \bibinfo{pages}{115--142}.
\newblock


\bibitem[\protect\citeauthoryear{Suh, Bentley, and Lottridge}{Suh
  et~al\mbox{.}}{2018}]%
        {suh2018s}
\bibfield{author}{\bibinfo{person}{Minhyang Suh}, \bibinfo{person}{Frank
  Bentley}, {and} \bibinfo{person}{Danielle Lottridge}.}
  \bibinfo{year}{2018}\natexlab{}.
\newblock \showarticletitle{" It's Kind of Boring Looking at Just the Face" How
  Teens Multitask During Mobile Videochat}.
\newblock \bibinfo{journal}{\emph{Proceedings of the ACM on Human-Computer
  Interaction}} \bibinfo{volume}{2}, \bibinfo{number}{CSCW}
  (\bibinfo{year}{2018}), \bibinfo{pages}{1--23}.
\newblock


\bibitem[\protect\citeauthoryear{Tian, Zhang, and Karger}{Tian
  et~al\mbox{.}}{2021}]%
        {tian2021system}
\bibfield{author}{\bibinfo{person}{Sunny Tian}, \bibinfo{person}{Amy~X Zhang},
  {and} \bibinfo{person}{David Karger}.} \bibinfo{year}{2021}\natexlab{}.
\newblock \showarticletitle{A System for Interleaving Discussion and
  Summarization in Online Collaboration}.
\newblock \bibinfo{journal}{\emph{Proceedings of the ACM on Human-Computer
  Interaction}} \bibinfo{volume}{4}, \bibinfo{number}{CSCW3}
  (\bibinfo{year}{2021}), \bibinfo{pages}{1--27}.
\newblock


\bibitem[\protect\citeauthoryear{Torre, Galluccio, and Coccoli}{Torre
  et~al\mbox{.}}{2022}]%
        {torre2022video}
\bibfield{author}{\bibinfo{person}{Ilaria Torre}, \bibinfo{person}{Ilenia
  Galluccio}, {and} \bibinfo{person}{Mauro Coccoli}.}
  \bibinfo{year}{2022}\natexlab{}.
\newblock \showarticletitle{Video augmentation to support video-based
  learning}. In \bibinfo{booktitle}{\emph{Proceedings of the 2022 International
  Conference on Advanced Visual Interfaces}}. \bibinfo{pages}{1--5}.
\newblock


\bibitem[\protect\citeauthoryear{Tu, Yuan, and Wang}{Tu et~al\mbox{.}}{2018}]%
        {tu2018you}
\bibfield{author}{\bibinfo{person}{Pei-Yun Tu}, \bibinfo{person}{Chien~Wen
  Yuan}, {and} \bibinfo{person}{Hao-Chuan Wang}.}
  \bibinfo{year}{2018}\natexlab{}.
\newblock \showarticletitle{Do you think what I think: Perceptions of delayed
  instant messages in computer-mediated communication of romantic relations}.
  In \bibinfo{booktitle}{\emph{Proceedings of the 2018 CHI Conference on Human
  Factors in Computing Systems}}. \bibinfo{pages}{1--11}.
\newblock


\bibitem[\protect\citeauthoryear{Ullmann, De~Liddo, and Bachler}{Ullmann
  et~al\mbox{.}}{2019}]%
        {ullmann2019visualisation}
\bibfield{author}{\bibinfo{person}{Thomas~Daniel Ullmann},
  \bibinfo{person}{Anna De~Liddo}, {and} \bibinfo{person}{Michelle Bachler}.}
  \bibinfo{year}{2019}\natexlab{}.
\newblock \showarticletitle{A Visualisation Dashboard for Contested Collective
  Intelligence. Learning Analytics to Improve Sensemaking of Group Discussion}.
\newblock \bibinfo{journal}{\emph{RIED: Revista Iboeroamericana de
  Educaci{\'o}n a Distancia (The Ibero-American Journal of Digital Education)}}
  \bibinfo{volume}{22}, \bibinfo{number}{1} (\bibinfo{year}{2019}),
  \bibinfo{pages}{41--80}.
\newblock


\bibitem[\protect\citeauthoryear{Wainfan and Davis}{Wainfan and Davis}{2004}]%
        {wainfan2004challenges}
\bibfield{author}{\bibinfo{person}{Lynne Wainfan} {and} \bibinfo{person}{Paul~K
  Davis}.} \bibinfo{year}{2004}\natexlab{}.
\newblock \bibinfo{booktitle}{\emph{Challenges in virtual collaboration:
  Videoconferencing, audioconferencing, and computer-mediated communications}}.
\newblock \bibinfo{publisher}{Rand Corporation}.
\newblock


\bibitem[\protect\citeauthoryear{Wang, Wen, and Rose}{Wang
  et~al\mbox{.}}{2017}]%
        {wang2017contrasting}
\bibfield{author}{\bibinfo{person}{Xu Wang}, \bibinfo{person}{Miaomiao Wen},
  {and} \bibinfo{person}{Carolyn Rose}.} \bibinfo{year}{2017}\natexlab{}.
\newblock \showarticletitle{Contrasting explicit and implicit support for
  transactive exchange in team oriented project based learning}.
\newblock \bibinfo{publisher}{Philadelphia, PA: International Society of the
  Learning Sciences.}
\newblock


\bibitem[\protect\citeauthoryear{Warner, Pavel, Nguyen, Agrawala, and
  Hartmann}{Warner et~al\mbox{.}}{2023}]%
        {warner2023slidespecs}
\bibfield{author}{\bibinfo{person}{Jeremy Warner}, \bibinfo{person}{Amy Pavel},
  \bibinfo{person}{Tonya Nguyen}, \bibinfo{person}{Maneesh Agrawala}, {and}
  \bibinfo{person}{Bj{\"o}rn Hartmann}.} \bibinfo{year}{2023}\natexlab{}.
\newblock \showarticletitle{SlideSpecs: Automatic and Interactive Presentation
  Feedback}.
\newblock  (\bibinfo{year}{2023}).
\newblock


\bibitem[\protect\citeauthoryear{Wiederhold}{Wiederhold}{2020}]%
        {wiederhold2020connecting}
\bibfield{author}{\bibinfo{person}{Brenda~K Wiederhold}.}
  \bibinfo{year}{2020}\natexlab{}.
\newblock \bibinfo{title}{Connecting through technology during the coronavirus
  disease 2019 pandemic: Avoiding “Zoom Fatigue”}.
\newblock
\newblock


\bibitem[\protect\citeauthoryear{Wohn, Peng, and Zytko}{Wohn
  et~al\mbox{.}}{2017}]%
        {wohn2017face}
\bibfield{author}{\bibinfo{person}{Donghee~Yvette Wohn}, \bibinfo{person}{Wei
  Peng}, {and} \bibinfo{person}{Doug Zytko}.} \bibinfo{year}{2017}\natexlab{}.
\newblock \showarticletitle{Face to face matters: communication modality,
  perceived social support, and psychological wellbeing}. In
  \bibinfo{booktitle}{\emph{Proceedings of the 2017 CHI Conference Extended
  Abstracts on Human Factors in Computing Systems}}.
  \bibinfo{pages}{3019--3026}.
\newblock


\bibitem[\protect\citeauthoryear{Xu, Tao, Feng, Raqui, and Ranwez}{Xu
  et~al\mbox{.}}{2021}]%
        {xu2021benchmarking}
\bibfield{author}{\bibinfo{person}{Binbin Xu}, \bibinfo{person}{Chongyang Tao},
  \bibinfo{person}{Zidu Feng}, \bibinfo{person}{Youssef Raqui}, {and}
  \bibinfo{person}{Sylvie Ranwez}.} \bibinfo{year}{2021}\natexlab{}.
\newblock \showarticletitle{A benchmarking on cloud based speech-to-text
  services for french speech and background noise effect}.
\newblock \bibinfo{journal}{\emph{arXiv preprint arXiv:2105.03409}}
  (\bibinfo{year}{2021}).
\newblock


\bibitem[\protect\citeauthoryear{Yadav, Shrivastava, Mohana~Prasad, Arsikere,
  Patil, Kumar, and Deshmukh}{Yadav et~al\mbox{.}}{2015}]%
        {yadav2015content}
\bibfield{author}{\bibinfo{person}{Kuldeep Yadav}, \bibinfo{person}{Kundan
  Shrivastava}, \bibinfo{person}{S Mohana~Prasad}, \bibinfo{person}{Harish
  Arsikere}, \bibinfo{person}{Sonal Patil}, \bibinfo{person}{Ranjeet Kumar},
  {and} \bibinfo{person}{Om Deshmukh}.} \bibinfo{year}{2015}\natexlab{}.
\newblock \showarticletitle{Content-driven multi-modal techniques for
  non-linear video navigation}. In \bibinfo{booktitle}{\emph{Proceedings of the
  20th international conference on intelligent user interfaces}}.
  \bibinfo{pages}{333--344}.
\newblock


\bibitem[\protect\citeauthoryear{Yang, Yim, Kim, and Shin}{Yang
  et~al\mbox{.}}{2022}]%
        {yang2022catchlive}
\bibfield{author}{\bibinfo{person}{Saelyne Yang}, \bibinfo{person}{Jisu Yim},
  \bibinfo{person}{Juho Kim}, {and} \bibinfo{person}{Hijung~Valentina Shin}.}
  \bibinfo{year}{2022}\natexlab{}.
\newblock \showarticletitle{CatchLive: Real-time Summarization of Live Streams
  with Stream Content and Interaction Data}. In \bibinfo{booktitle}{\emph{CHI
  Conference on Human Factors in Computing Systems}}. \bibinfo{pages}{1--20}.
\newblock


\bibitem[\protect\citeauthoryear{Zhang and Cranshaw}{Zhang and
  Cranshaw}{2018}]%
        {zhang2018making}
\bibfield{author}{\bibinfo{person}{Amy~X Zhang} {and} \bibinfo{person}{Justin
  Cranshaw}.} \bibinfo{year}{2018}\natexlab{}.
\newblock \showarticletitle{Making sense of group chat through collaborative
  tagging and summarization}.
\newblock \bibinfo{journal}{\emph{Proceedings of the ACM on Human-Computer
  Interaction}} \bibinfo{volume}{2}, \bibinfo{number}{CSCW}
  (\bibinfo{year}{2018}), \bibinfo{pages}{1--27}.
\newblock


\bibitem[\protect\citeauthoryear{Zhang, Verou, and Karger}{Zhang
  et~al\mbox{.}}{2017}]%
        {zhang2017wikum}
\bibfield{author}{\bibinfo{person}{Amy~X Zhang}, \bibinfo{person}{Lea Verou},
  {and} \bibinfo{person}{David Karger}.} \bibinfo{year}{2017}\natexlab{}.
\newblock \showarticletitle{Wikum: Bridging discussion forums and wikis using
  recursive summarization}. In \bibinfo{booktitle}{\emph{Proceedings of the
  2017 ACM Conference on Computer Supported Cooperative Work and Social
  Computing}}. \bibinfo{pages}{2082--2096}.
\newblock


\bibitem[\protect\citeauthoryear{Zhang, Tao, Chen, Sun, Judson, and
  Naqvi}{Zhang et~al\mbox{.}}{2018}]%
        {zhang2018co}
\bibfield{author}{\bibinfo{person}{Jianwei Zhang}, \bibinfo{person}{Dan Tao},
  \bibinfo{person}{Mei-Hwa Chen}, \bibinfo{person}{Yanqing Sun},
  \bibinfo{person}{Darlene Judson}, {and} \bibinfo{person}{Sarah Naqvi}.}
  \bibinfo{year}{2018}\natexlab{}.
\newblock \showarticletitle{Co-organizing the collective journey of inquiry
  with idea thread mapper}.
\newblock \bibinfo{journal}{\emph{Journal of the Learning Sciences}}
  \bibinfo{volume}{27}, \bibinfo{number}{3} (\bibinfo{year}{2018}),
  \bibinfo{pages}{390--430}.
\newblock


\end{thebibliography}
\received{January 2023}
\received[revised]{April 2023}
\received[accepted]{May 2023}

\appendix
\section{Appendix}
\begin{table}[h]
\small
{%
\begin{tabular}{llllll}
\toprule
Gender & Age               & Ethnicity        & Location      & Education         & Occupation         \\ \midrule
Male   & 21 - 30  & Asian            & Asia          & Bachelor's Degree & Graduate Student   \\
Female & 10 - 20  & Asian            & Asia          & High School       & College Student    \\
Female & 21 - 30 & Asian            & North America & Bachelor's Degree & Graduate Student   \\
Male   & 30 - 40  & White            & North America & Ph.D. or higher   & Professor          \\
Female & 30 - 40  & Asian            & North America & Ph.D. or higher   & Professor          \\
Female & 21 - 30  & Asian            & Asia          & Bachelor's Degree & Graduate Student   \\
Male   & 21 - 30  & Asian            & Asia          & Bachelor's Degree & Product Manager    \\
Female & 21 - 30 & Asian            & Europe        & High School       & College Student    \\
Male   & 30 - 40 & Asian            & Asia          & Bachelor's Degree & Engineer           \\
Female & 10 - 20 & Asian            & Asia          & High School       & College Student    \\
Male   & 21 - 30  & Black or African American           & North America & High School       & College Student    \\
Male   & 21 - 30 & White            & Europe        & Bachelor's Degree & Programmer         \\
Female & 21 - 30 & Black or African American  & North America & High School       & E-commerce        \\
Male   & 21 - 30 & Black or African American           & Africa        & Bachelor's Degree & Pharmacist         \\
Female & 10 - 20 & Black or African American   & North America & High School       & E-commerce        \\
Male   & 21 - 30  & Black or African American   & North America & Bachelor's Degree & Software Engineer  \\
Male   & 50+     & White   & North America & Ph.D. or higher   & High School Principal          \\
Female & 50+     & Black or African American & North America & Master's Degree   & program consultant \\
Male   & 21 - 30  & Asian            & North America & Bachelor's Degree & Game Developer     \\
Female & 10 - 20  & White            & North America & Bachelor's Degree & Graduate Student   \\
Female & 21 - 30  & White            & Europe        & Bachelor's Degree & Marketing Professionals \\        \bottomrule    
\end{tabular}%
}
\caption{The demographic information of participants in the speed-dating study}
\label{demo}
\end{table}

\begin{table}[h]
\small
\centering
\begin{tabular}{p{4cm}p{3cm}p{0.5cm}p{6cm}}
\hline
\textbf{Scenario} & \textbf{Design Concept} & & \textbf{Solution in Storyboards} \\ \hline
The listener has difficulties in comprehension
& Provide ways for users to keep track of the discussion  & 1 & Provide real-time interactive transcripts to which people can collaboratively add annotations.\\   \hline
The speaker has difficulties in conveying ideas & Provide more channels to convey ideas & 2 & Users can sketch ideas and append it to the corresponding transcript \\ \hline
\multirow{2}{4cm}{The speaker receives misleading feedback from the audience} & \multirow{2}{3cm}{Provide explicit way for the speaker to check understanding} & 3 & The speaker can send an understanding check and questions are automatically extracted from transcript  \\ \hline
\multirow{2}{4cm}{When someone is confused, they lack channels to express confusion comfortably} & \multirow{2}{3cm}{Provide more ways for people to give feedback} &  4 & Enable anonymous and private comments on transcripts for people to ask questions  \\ \cline{3-4} 
&& 5 & Users could mark sentences in the live transcript as "confusing", and the speaker can receive notifications \\ \hline
Some people are shy and cannot find time to speak or ask questions & Give everyone opportunities to express ideas & 6 & Users can participate through text/reactions/emojis on the transcript.\\ \hline
\multirow{2}{4cm}{Users feel cognitively occupied when balancing listening, speaking, and managing the process at the same time} & Reduce the overload of reading full transcript & 7 & Only present important transcript to users \\ \cline{2-4} 
 & Include time management services in meetings & 8 & Participants can upload their agenda and time allocation plan and see a progress bar during the meeting.\\ \hline
\multirow{2}{4cm}{People cannot process the information when others speak for too long} & Reduce the efforts on processing and synthesizing information & 9 & Provide summaries and key points\\ \cline{2-4}
& Extract highlights from monologues & 10 & The speakers themselves can highlight what is important in the transcript when speaking \\ \hline
\end{tabular}%
\caption{\textbf{The list of scenarios, design concepts, and storyboards.} The goal of the speed-dating study is to quickly test design ideas and explore participants' preferences and boundaries when using technologies to support active participation in group video meetings.}
\label{tab:concept-storyboard}
\end{table}

\begin{table}[]
\resizebox{\columnwidth}{!}{%
\begin{tabular}{p{2cm}|p{3.5cm}|p{4cm}|p{3.5cm}|p{4cm}}
\hline
\multirow{2}{*}{} &
  \multicolumn{2}{l}{\textbf{Otter.ai}} &
  \multicolumn{2}{l}{\textbf{MeetScript}} \\ \cline{2-5} 
 &
 \textbf{Features} &
\textbf{Pros \& Cons} &
 \textbf{Features} &
  \textbf{Pros \& Cons }\\ \hline
\multirow{4}{2cm}{\textbf{Transcript Interface}} &
  document-like interface &
  low readability &
  chat message with different color &
  more readable \\ \cline{2-5}
 &
  Output transcript in a long paragraph &
  Hard to find information & 
  Chunk transcript into bubbles based on single utterance &
  Less information density \\ \cline{2-5}
 &
  Automatic speaker recognition post-meeting &
  No in-meeting speaker differentiation services. Not stable. &
  Differentiate speakers through browser input Conversation &
  More stable \\ \cline{2-5}
 &
  The transcript is separated with the video conferencing interface &
  more configuration to connect the service &
  Transcript is connected to the audio/video in the same interface &
  Less configuration effort \\ \hline
\multirow{7}{2cm}{\textbf{Features for in-situ participation}} &
  Edit in post-meeting &
  Can not correct transcript errors during meeting &
  Edit in real-time and post-meeting &
  Enable add/delete/edit transcript during meeting \\ \cline{2-5}
 &
  N/A &
  N/A &
  Tag to add high-level label &
  Help with synthesize and re-organize the discussion \\ \cline{2-5}
 &
  Comment based on selection and integrated into the "Takeaway’’ panel &
  Comment is separated from the transcript and integrated to a side panel. It is good to synthesize takeaways but not proper for real-time contribution to the conversion. &
  The comment is based on transcript bubbles &
  People can participate in the conversation through the threaded comment. And they can synthesize the comment in the interaction Interaction History Panel. The anonymous transcript is enabled \\ \cline{2-5}
 &
  Highlight &
  Only one color &
  Highlight &
  In different colors \\ \cline{2-5}
 &
  N/A &
  N/A &
  like &
  Quick archive and attitude expression \\ \cline{2-5}
 &
  Assign action item &
  Useful for task assignment &
  N/A &
  Can be realized through tagging \\ \cline{2-5}
 &
  Insert screenshot &
  Useful for capturing image information &
  N/A &
  N/A \\ \hline
\multirow{4}{2cm}{\textbf{Features to help keep up with the conversation}} &
  All the transcript is kept on the interface &
  Hard to quickly retrieve the long transcript &
  Only transcript with interaction is left on the interface &
  More easily to find information with only informative transcript \\ \cline{2-5}
 &
  Takeaway panel to show all interactions &
  Helpful for synthesizing and locating information &
  Interaction Interaction History Panel to show one’s own interaction &
  Helpful for synthesizing and locating information \\ \cline{2-5}
 &
  N/A &
  N/A &
  Minimap to visualize the information density &
  Quickly navigate the long transcript during the discussion \\ \cline{2-5}
 &
  AI-generated outline/keywords in post-meeting &
  Can be useful to review the conversation post-meeting but not accurate enough &
  N/A &
  N/A \\ \bottomrule
\end{tabular}%
}
\caption{The novelty of the MeetScript system}
\label{tab:my-table}
\end{table}

\begin{figure}[h] 
\centering 
\includegraphics[width=\textwidth]{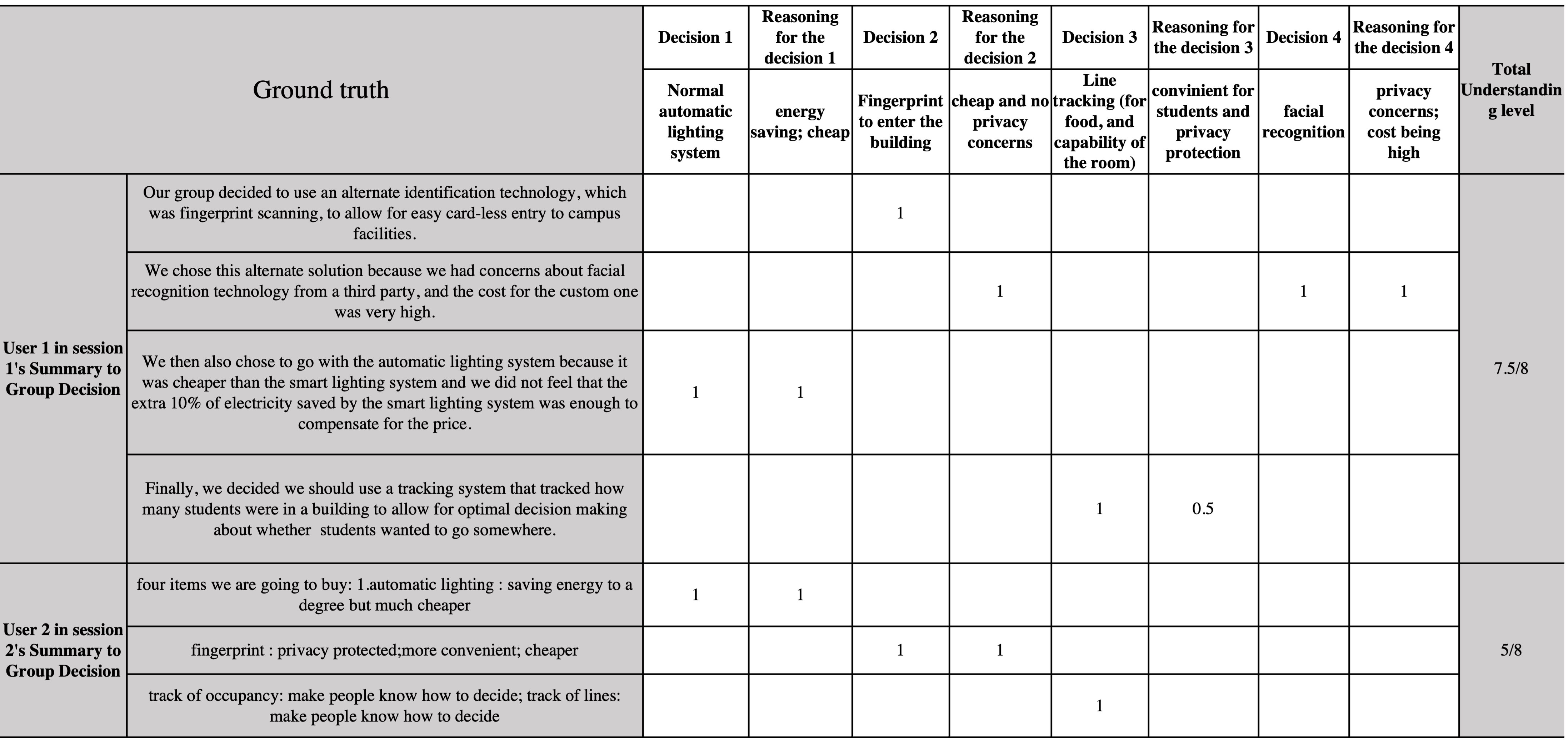} 
\caption{\textbf{Example coding for measuring the understanding level.} One author generated ground truth answers for the final decision and decision-making process of a group. Two authors then coded all participants' answers, assigning them a score based on the products chosen and the reasoning behind those choices. They calculated a normalized understanding level by using the ratio of each participant's final score to the group's total score. }
\label{coding}
\end{figure}

\end{document}